\documentclass[twocolumn,nofootinbib,showpacs,prd]{revtex4}

\usepackage{graphics,graphicx}

\begin{document}

\title{Where post-Newtonian and numerical-relativity waveforms meet}

\author{Mark Hannam$^1$, Sascha Husa$^1$, Jos\'e A. Gonz\'alez$^{1,2}$, Ulrich
  Sperhake$^1$, Bernd Br\"ugmann$^1$} 
\address{1 Theoretical Physics Institute, University of Jena, 07743 Jena, Germany}
\address{2 Instituto de F\a'{\i}sica y Matem\a'aticas,
Universidad Michoacana de San Nicol\a'as de Hidalgo, Edificio C-3,
Cd. Universitaria. C. P. 58040 Morelia, Michoac\a'an, M\a'exico}

\begin{abstract}
We analyze numerical-relativity (NR) waveforms that cover nine orbits
(18 gravitational-wave cycles) before merger of an equal-mass system with low
eccentricity, with  
numerical uncertainties of 0.25 radians in the phase
and less than 2\% in the amplitude; such accuracy allows a
direct comparison with post-Newtonian (PN) waveforms. 
We focus on one of the PN approximants that has been proposed for use in
gravitational-wave data analysis, the restricted 3.5PN ``TaylorT1'' waveforms,
and compare these 
with a section of the numerical waveform from the second to the eighth orbit,
which is about one and a half orbits before merger. This corresponds to a gravitational-wave
frequency range of $M\omega = 0.0455$ to $0.1$. Depending on
the method of matching PN and NR waveforms, the accumulated phase disagreement
over this frequency range can be within numerical uncertainty. Similar results
are found in comparisons with an alternative PN  approximant, 3PN
``TaylorT3''.
The amplitude disagreement, on the other hand, is around 6\%, but roughly
constant for all 13 cycles that are compared, suggesting that only 4.5 orbits
need be simulated to match PN and NR waves with the same accuracy as is
possible with nine orbits. If, however, we model the amplitude up to 2.5PN
order, the amplitude disagreement is roughly within numerical uncertainty up
to about 11 cycles before merger.

\end{abstract}


\maketitle

\section{Introduction}

The current generation of interferometric gravitational-wave detectors 
~\cite{Waldman06,GEOStatus:2006,Acernese2006} 
have reached design sensitivity, and have recently completed taking data in the S5 science 
run. Signals from coalescing
black-hole binaries will be among the strongest  
that one hopes to find in the detector data, and data analysts are searching for
them by performing matched filtering against template banks of theoretical 
waveforms. At present data analysts use, or are preparing to use, waveforms
calculated by post-Newtonian (PN) methods, and in particular the standard
Taylor-expanded,
effective-one-body (EOB) and BCV \cite{Buonanno03a} 
waveforms implemented in the LSC Algorithms Library (LAL) \cite{LAL}, although
current GW searches do not go beyond second PN order (2PN).  The PN
waveforms are expected to be reasonably accurate during the slow inspiral of the 
binaries, but it is not clear how well they can model the merger phase.
Ultimately the PN waveforms will be connected to waveforms
from fully general relativistic numerical simulations, which will also model
the last orbits, merger and ringdown. 

In the last two years breakthroughs in numerical relativity \cite{Pretorius:2005gq,
Campanelli:2005dd,Baker05a} have completed the work of providing the
techniques to generate the 
necessary numerical (NR) waveforms. The nonspinning equal-mass case in
particular has been studied in great detail 
\cite{
Pretorius:2005gq,Pretorius:2006tp,Campanelli:2005dd,Baker05a,Campanelli:2006gf,
Baker:2006yw,Sperhake2006,Bruegmann:2006at,Baker:2006ls,Baker:2006ha,
Pfeiffer:2007,Scheel-etal-2006:dual-frame} 
and extremely accurate waveforms over
many ($>15$) cycles are now available \cite{Husa2007a}. A first comparison of PN 
and NR equal-mass waveforms was made in
\cite{Buonanno06imr,Baker:2006ha}, unequal-mass waveforms were studied
in \cite{Berti:2007fi,Pan:2007nw,Buonanno:2007pf}, and spinning binaries in
\cite{Vaishnav:2007nm}, and the work of producing hybrid NR-PN waveforms has 
begun \cite{Pan:2007nw,Ajith:2007qp,Ajith:2007kx,Buonanno:2007pf}. Good agreement has been observed
between NR and PN waveforms \cite{Buonanno06imr,Baker:2006ha,Pan:2007nw}, and
in particular phase disagreements of less than one radian up to $\sim 1.5$
orbits before merger were seen in \cite{Baker:2006ha}. However, until now
NR waveforms have not been accurate enough to allow a conclusive comparison
with the PN wave amplitude; for example, it was pointed out in
\cite{Pan:2007nw} that although the 
disagreement in the amplitude of NR and PN waves was about 10\%, this was also
the size of the uncertainty in the NR wave amplitude, and it was not possible
to conclude what order of PN treatment of the wave amplitude gives the best
agreement with fully general-relativistic results.

In this work we systematically compare numerical equal-mass waveforms that 
include up to 18 cycles before merger with the 3.5PN ``TaylorT1'' and 3PN ``TaylorT3''
waveforms implemented in LAL. One could compare with many different varieties
of PN waveform, but the T1 and T3 approximants are common choices that are
among those proposed for gravitational-wave searches in detector data, and
restricting ourselves to only two approximants keeps our analysis and the
presentation of our results relatively simple. The region of comparison includes 13
cycles. Considering the amplitude $A(t)$ and phase $\phi(t)$ of our 
numerical waveforms separately, we find that the accumulated error in
$\phi(t)$ is at most 0.25 radians over the frequency range of
comparison. These uncertainties are dominated by the finite extraction radii of our
waveforms, {\it not} finite-difference errors. The
error in the amplitude $A(t)$ is less than 2\% for most of the simulation. 
We estimate the eccentricity as $e < 0.0016$. We therefore consider 
these waveforms to be adequately accurate for a detailed comparison with PN results,
in particular to determine the disagreement between NR and PN wave amplitudes. 

Numerical simulations are computationally expensive, and mapping the parameter 
space of binary mergers (including black holes of varying mass ratio and spins) will 
require huge computer resources. As such, we would prefer to simulate only a small 
number of orbits before matching with PN results. We find that a simulation of only 
4.5 orbits has the same amplitude agreement with the last four cycles of
the restricted 3.5PN waveform as the long 18-cycle simulation, and therefore  
suggest that relatively short numerical simulations are feasible for matching to 
PN inspiral waves. For an even greater amplitude agreement with PN theory, our
results suggest that, using 2.5PN amplitude corrections, at least 5.5 orbits
(11 cycles) before merger are necessary. 

We also compare the black holes' motion with that calculated by integrating 
 the PN equations of motion \cite{Buonanno:2005xu,Husa:2007ec} and find that
 the PN and NR orbital tracks and frequencies are in excellent agreement until
 the last three orbits of the binary. 
 
Before describing our analysis in detail, we give a brief summary of our
numerical methods in Section~\ref{sec:numerical} and the procedure for
generating PN waveforms in Section~\ref{sec:pn}. In
Section~\ref{sec:simulations} we discuss the simulations we 
performed, and in particular establish the sixth-order convergence of our
results, construct Richardson extrapolated waveforms with error estimates, and
extrapolate the finite-extraction-radius waveforms to those measured as
$R_{ex} \rightarrow \infty$. We also discuss the phase errors in our waveforms
and give a consistency check between waves from simulations starting at 
different initial separations. In Section~\ref{sec:comparison} we directly
compare the PN and numerical waveforms.

\section{Numerical methods and waveforms}
\label{sec:numerical}

We performed numerical simulations with the BAM code
\cite{Bruegmann:2006at,Bruegmann2004}, replacing fourth-order accurate
derivative operators by sixth-order accurate spatial derivative operators in the bulk 
as described in
\cite{Husa2007a}. The code starts with black-hole binary puncture initial data 
\cite{Brandt97b,Bowen80} generated using a pseudo-spectral code
\cite{Ansorg:2004ds}, and evolves them with the $\chi$-variant of the
moving-puncture \cite{Campanelli2006,Baker2006} version of the BSSN
\cite{Shibata95,Baumgarte99} formulation of the 3+1 Einstein 
evolution equations \cite{York79}. 
The gravitational waves emitted by the binary are calculated from the
Newman-Penrose scalar $\Psi_4$, and the details of our implementation of
this procedure are given in \cite{Bruegmann:2006at}.

The simulations we performed for this analysis are summarized in 
Tables~\ref{tab:simulations} and \ref{tab:parameters}. For the configurations
with initial separations  
$D = 10M, 11M, 12M$ (denoted by ``D10'', ``D11'' and ``D12'' throughout the
paper), simulations were performed at three resolutions, and
final results obtained by Richardson extrapolation, as described in 
Section~\ref{sec:simulations}. For the $D=8M, 9M$ (``D8'', ``D9'') configurations, 
which are used only for comparison at the end of Section~\ref{sec:comparison},
only one simulation at medium resolution was performed. 

\begin{table}
\caption{\label{tab:simulations}
Summary of grid setup for numerical simulations.
} 
\begin{tabular}{||l|r|r|r|}
\hline
 Grid & $h_{min}$ & $h_{max}$ & $r_{max}$ \\
\hline
\multicolumn{4}{||l|}{ D8 simulation } \\
\hline
$\chi_{\eta=2}[5\times 56:5\times 112:6] $ & $M/37.3$ &  $96/7 M$ &
$775M $\\
\hline
\multicolumn{4}{||l|}{D9 simulation }\\
\hline
$\chi_{\eta=2}[5\times 56:5\times 112:6] $ & $M/37.3$ &  $96/7 M$ &
$775M$ \\
 \hline
\multicolumn{4}{||l|}{ D10 and D11 simulations} \\
 \hline
$\chi_{\eta=2}[5\times 48:5\times  96:6] $ & $M/32.0$ &  $16 M$   & $776M$ \\
$\chi_{\eta=2}[5\times 56:5\times 112:6] $ & $M/37.3$ &  $96/7 M$ & $775M$\\
$\chi_{\eta=2}[5\times 64:5\times 128:6] $ & $M/42.7$ &  $12 M$  & $774M$\\
\hline
\multicolumn{4}{||l|}{D12 simulations }\\
\hline
$\chi_{\eta=2}[5\times 64:5\times 128:6] $ & $M/42.7$ &  $12 M$ & $774M$ \\
$\chi_{\eta=2}[5\times 72:5\times 144:6] $ & $M/48.0$ &  $32/3 M$ & $773M$ \\
$\chi_{\eta=2}[5\times 80:5\times 160:6] $ & $M/53.3$ &  $48/5 M$  & $773M$ \\
\hline
\end{tabular}
\end{table}

The physical parameters are given in Table~\ref{tab:parameters}. The initial
momenta for low-eccentricity quasi-circular inspiral are estimated by the PN
method described in \cite{Husa:2007ec}. We estimated
the eccentricity from the frequency $\omega_p$ of the puncture 
motion, as we did previously for $D = 11M$ simulations in \cite{Husa:2007ec}, 
and as also used in \cite{Buonanno06imr,Baker:2006ha}. Given the puncture
motion frequency $\omega_p(t)$ and the frequency of a comparable 
zero-eccentricity simulation $\omega_c (t)$ (estimated by fitting a fourth-order
polynomial in $t$ through the numerical $\omega_p(t)$, as suggested in
\cite{Baker:2006ha}), the eccentricity is estimated by finding extrema in
the function $(\omega_p(t) - \omega_c(t))/(2 \omega_c(t))$. The uncertainty in
the eccentricity estimate is about $2 \times 10^{-4}$ \cite{Husa:2007ec}. A simulation with
initial $D = 12M$ but using ``quasi-circular orbit'' 
parameters (as discussed further in Section~\ref{sec:eccentricity}) has 
an eccentricity of $e \approx 0.008$, i.e., five times larger than the
eccentricity of the D12 simulation. 

\begin{table}
\caption{\label{tab:parameters}
Physical parameters for the moving-puncture simulations: the coordinate
separation, $D/M$, the mass parameters in the puncture data construction,
$m_i/M$, and the momenta $p_x/M$ and $p_y/M$. The momenta are based on those
described in \cite{Husa:2007ec}, and produce quasi-circular inspiral with
minimal eccentricity. The estimated eccentricity $e$ is also given, as
described in the text.  The punctures are placed on the $y$-axis, and for all
simulations the total initial black-hole mass is $M = 1$.} 
\begin{tabular}{||l|r|r|r|r|r|}
\hline
Simulation & $D/M$ & $m_i/M$ & $p_x/M$        & $p_y/M$ ($\times 10^{-3}$) & $e$ \\
\hline 
D8         & 8.0   & 0.48240 & $\mp 0.11235$  & $\mp 2.0883$       & $0.0025$ \\
D9         & 9.0   & 0.48436 & $\mp 0.10337$  & $\mp 1.4019$       & $0.0022$  \\
D10        & 10.0  & 0.48593 & $\mp 0.096107$ & $\mp 0.980376$     & $0.0022$ \\
D11        & 11.0  & 0.48721 & $\mp 0.090099$ & $\mp 0.709412$     & $0.0020$ \\
D12        & 12.0  & 0.48828 & $\mp 0.085035$ & $\mp 0.537285$     & $0.0016$ \\
\hline
\end{tabular}
\end{table}

For comparison between PN and numerical results, we must make clear what we
mean by the individual black hole masses $M_1$ and $M_2$, and the total mass
$M$. The mass of each black hole, $M_i$, is specified 
in terms of the Arnowitt-Deser-Misner (ADM) 
mass at each puncture. This corresponds to the mass at the other
asymptotically flat end, and has been found to equal numerically the
apparent-horizon mass \cite{Tichy:2003qi}, which for nonspinning black holes
is related to the area of the horizon $A_i$ by 
\begin{equation} 
M_i = \sqrt{ \frac{A_i}{16 \pi} }.
\end{equation}
We assume that this mass is the same as the mass used in post-Newtonian formulas. 
Rather than try to quantify the accuracy of this assumption, we make the
following argument, which we consider to be a more practical approach.
Our assumption is rigorously true only in the limit where the black holes are
infinitely far apart and stationary. As such we consider any error in this
assumption as part of the error due to starting the simulation at a finite
separation. Since there is no invariant measure of quasi-local mass in general
relativity, this error is present in some form in all numerical
simulations. In practice one could rescale the total mass to optimize the
phase and amplitude agreement with post-Newtonian calculations, but in the
present work we retain the assumption that the horizon mass and PN masses can
be equated. This provides an overall scale $M = M_1 + M_2$ for both numerical
and post-Newtonian waveforms, and is crucial for comparison and matching. 

Let us discuss some other possible sources of error related to the masses and
spins of the black holes in our numerical simulations. 

The initial data contain ``junk'' radiation that quickly leaves
the system. Some of this radiation may fall into the black holes and alter
their masses. To estimate this effect, we refer to the initial-data studies of
Cook and York \cite{Cook90a,Cook90}, who estimated the maximum radiation
content of single boosted Bowen-York black-hole initial-data sets (recall that
a single boosted Schwarzschild black-hole spacetime will not contain any
gravitational radiation). An estimate based on their data suggests that the
spacetime of a Bowen-York black-hole with $P_i/M_i \approx 0.17$ (which is the
case for the D12 simulations) has a maximum gravitational-wave energy content
of 0.01\% of the mass. In our simulations, the radiated energy from the junk
radiation is at least 0.005\% of the initial mass. Therefore we estimate that
at most 0.005\% of the mass fell back into the black hole. An error in our
estimate of the total mass of 0.005\% would lead to a phase error in a 2000$M$
simulation of 0.1$M$. We calculate (See Section~\ref{sec:simulations}) a
numerical uncertainty in the merger time of 0.4$M$, making any effect due to
junk radiation falling into the black holes lower than our numerical
uncertainty, and therefore not detectable at our level of accuracy. 

A further possible issue with the mass is that it may drift due to numerical
error over the course of the simulation. However, since we see clean
sixth-order convergence in the time when the gravitational wave reaches a
maximum, we expect that any mass drift either also converges away at
sixth-order, or is well below the error due to other numerical effects. 

Finally, one may worry that the black holes pick up spin during their
evolution. This effect has already been studied by Campanelli, {\it et. al.}
\cite{Campanelli:2006vp}. We do not attempt to measure this effect in our
simulations, for the following reason: we are comparing numerical and PN
waveforms of binaries that initially consist of nonspinning black holes. In
the PN approach we use, the black holes remain nonspinning. Any spin that they
acquire in full general relativity will therefore contribute to the
disagreement between PN and NR waveforms. It is that difference in the
waveforms that we are interested in measuring. More detailed investigation of
the physical properties of nonspinning binaries is beyond the scope of this
study.

\section{post-Newtonian waveforms}
\label{sec:pn}

Binary inspiral waveforms can be constructed by a variety of means.
We choose to compare our 
numerical waveforms with particular PN waveforms that are proposed for future 
searches for gravitational wave signals from black hole binary coalescence, 
namely the Taylor-expanded or EOB-resummed waveforms implemented in the LSC 
Applications Library (LAL) \cite{LAL,Damour00a,Damour:2002kr}. In particular, 
we compare with the 
3.5PN Taylor T1 waves, with a version of the
code\footnote{
We used a version of LAL consistent with cvs version 1.25; earlier versions contain
errors in the TaylorT1 implementation.}
that includes
modifications to the flux coefficients given in the Erratum to
\cite{Damour00a,Damour:2002kr}. 
In the Taylor T1 approach ordinary differential equations are solved numerically
to give the phase of the wave, and the amplitude is estimated by the quadrupole
contribution, which is proportional to $x=(M\omega/2)^{2/3}$, where $\omega$ is
the gravitational-wave frequency, and $\omega/2$ is assumed to be the orbital 
frequency of the binary. This treatment of the
amplitude yields ``restricted'' PN waveforms. In Section~\ref{sec:comparison} we
also compare with a 2.5PN treatment of the amplitude \cite{Arun04}, which 
includes terms up to $x^{7/2}$. 

To check the consistency of our comparison, we also compare with the ``Taylor
T3'' PN approximant \cite{Blanchet:2004ek,Blanchet:2001ax}, which 
consists of an analytic expression for the gravitational-wave phase as a
function of the variable $\tau = \nu (t - t_c) / (5M)$, where $t_c$ is the
``coalescence time'' 
of the binary, $M$ is the total mass, and $\eta = M_1 M_2 / M^2$ is the symmetric
mass ratio. The T3 approximant for the phase also contains a free phase
constant, $\phi_0$. The coalescence time $t_c$ and phase constant $\phi_0$ can
be chosen to line up the phase and frequency of a T3 PN waveform with an NR
waveform at a given time. We use the TaylorT3 approximant up to 3PN order,
because the 3.5PN term contains an unphysical turning point long before the
merger, which was already noted in the PN comparisons made in
\cite{Buonanno06imr,Baker:2006ha}. 

The LAL code that we use, {\tt LALInspiralTest}, produces $h_+$ and/or $h_-$
as a function of time. From this function we can compute  
the real part of $\Psi_{4,(l=2,m=2)}$ by differentiating twice with respect to
time. We choose $\Psi_{4,22}$ as
the quantity to compare between NR and PN waveforms for two reasons: (1)
we can compute the PN $h_{+,-}$ with arbitrarily small discretization error,
and thus expect that its derivatives
will be more accurate than computing $h_{+,-}$ by integration of the NR
$\Psi_{4,22}$; (2) integration of 
$\Psi_{4,22}$ requires estimating two constants of integration, which further 
complicates the procedure. In short, it should be equivalent to compare
waveforms using $h_{+,-}$ and $\Psi_{4,22}$, and we choose $\Psi_{4,22}$
because it is more straightforward.  

To generate a PN waveform we must choose
the masses of the two bodies, and a range of frequencies that we want the
waveform to cover. The masses are specified in units of solar mass. 
To produce the quantity $r \Psi_{4,22}$ that we wish to compare with numerical
data, the time is rescaled to be in units of $M$ by multiplying by the factor
$c/M$, where we chose $M = 2M_{\odot} = 2953.25$~m (although the choice of
masses is arbitrary) and the speed of light is
$c = 2.9979 \times 10^8$ m/s. The 3.5PN wave strain is then
differentiated twice with respect to time, and the amplitude is scaled by the
factor $\sqrt{16 \pi/5}$ to give the coefficient of the $l=2,m=2$ mode.

\section{Numerical simulations: accuracy and consistency}
\label{sec:simulations}

In this section we describe our procedure for producing the most accurate
waveform possible from our numerical data. This consists of taking waveforms
calculated at five extraction radii from simulations performed at three
resolutions, and (1) Richardson extrapolating these waveforms with respect to
numerical resolution to produce accurate waveforms at each of the five
extraction radii, and then (2) extrapolating with respect to extraction radius
to estimate the signal that would be calculated as $R_{ex} \rightarrow
\infty$. 

In \cite{Husa2007a} we described the use of sixth-order accurate spatial
finite differencing in the bulk in BAM, introduced to increase the overall
accuracy and in particular reduce the phase error in long evolutions. 
We found that $Re(r \Psi_4)_{22}$, as directly computed by the code, was
sixth-order convergent only up to about $100M$ before merger. However, if we
separate the waveform 
into its amplitude and frequency as \begin{equation}
r \Psi_4 = A(\phi(t)) e^{i\phi(t)}, \label{eqn:AmpPhase}
\end{equation} and examine separately $\phi(t)$ and $A(\phi)$, then
the phase shows reasonably clean sixth-order convergence throughout the
evolution (with a small ``blip'' around the merger time), and the amplitude
computed as a function of the phase angle shows good convergence with far
lower errors than when we consider simply $A(t)$. This is because $A(t)$
includes errors from the phase as well as the amplitude measurement;
considering $A(\phi)$ allows us to isolate the phase errors from the amplitude
errors. With this phase/amplitude split we are able to perform Richardson
extrapolation and reconstruct a more accurate waveform and calculate an error
estimate. More details about the convergence properties of these simulations
can be found in \cite{Husa2007a}.

In order to be as clear as possible about this procedure, we
will outline in detail the steps we followed to produce the D12 waveform that
will form the basis of our comparison with PN waveforms. 

We perform three simulations with the grid configuration (following the
notation in \cite{Bruegmann:2006at}) $\chi_{\eta=2}[5\times N:5\times  2N:6]$,
where $N = 64,72,80$ for the D12 runs. The grid resolutions on the finest
inner box are $M/42.67$, $M/48$ and $M/53.33$, and the resolutions on the
coarsest outer levels are $12M$, $10.67M$ and $9.6M$, placing the outer
boundary at about $775M$. The wave extraction is
performed at resolutions $1.5M$, $1.33M$ and $1.2M$. The grid
setup is summarized in Table~\ref{tab:simulations}, which also provides
the grid details for D8, D9, D10 and D11 simulations. 

\begin{figure}[t]
\centering
\includegraphics[height=4cm]{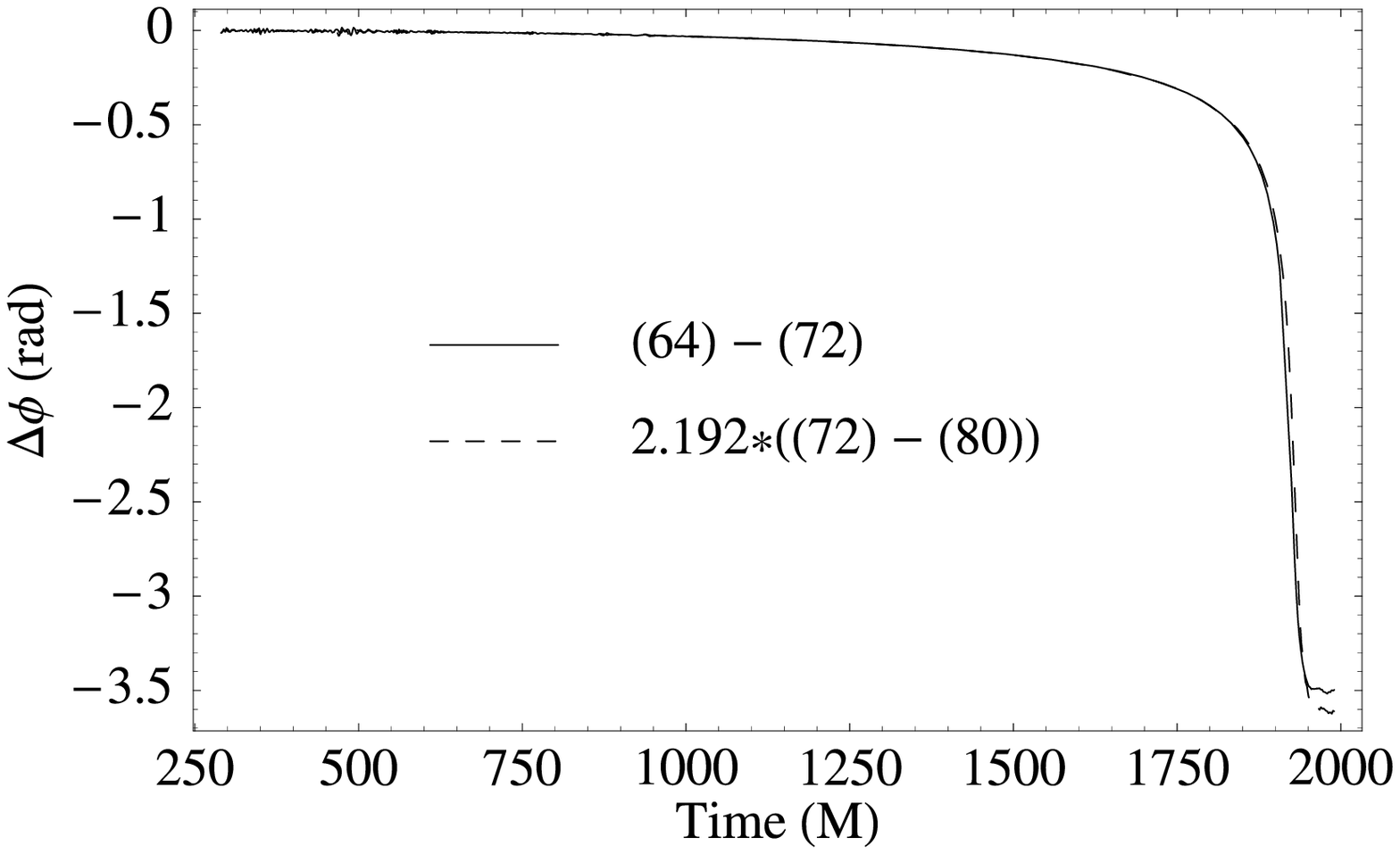}
\includegraphics[height=4cm]{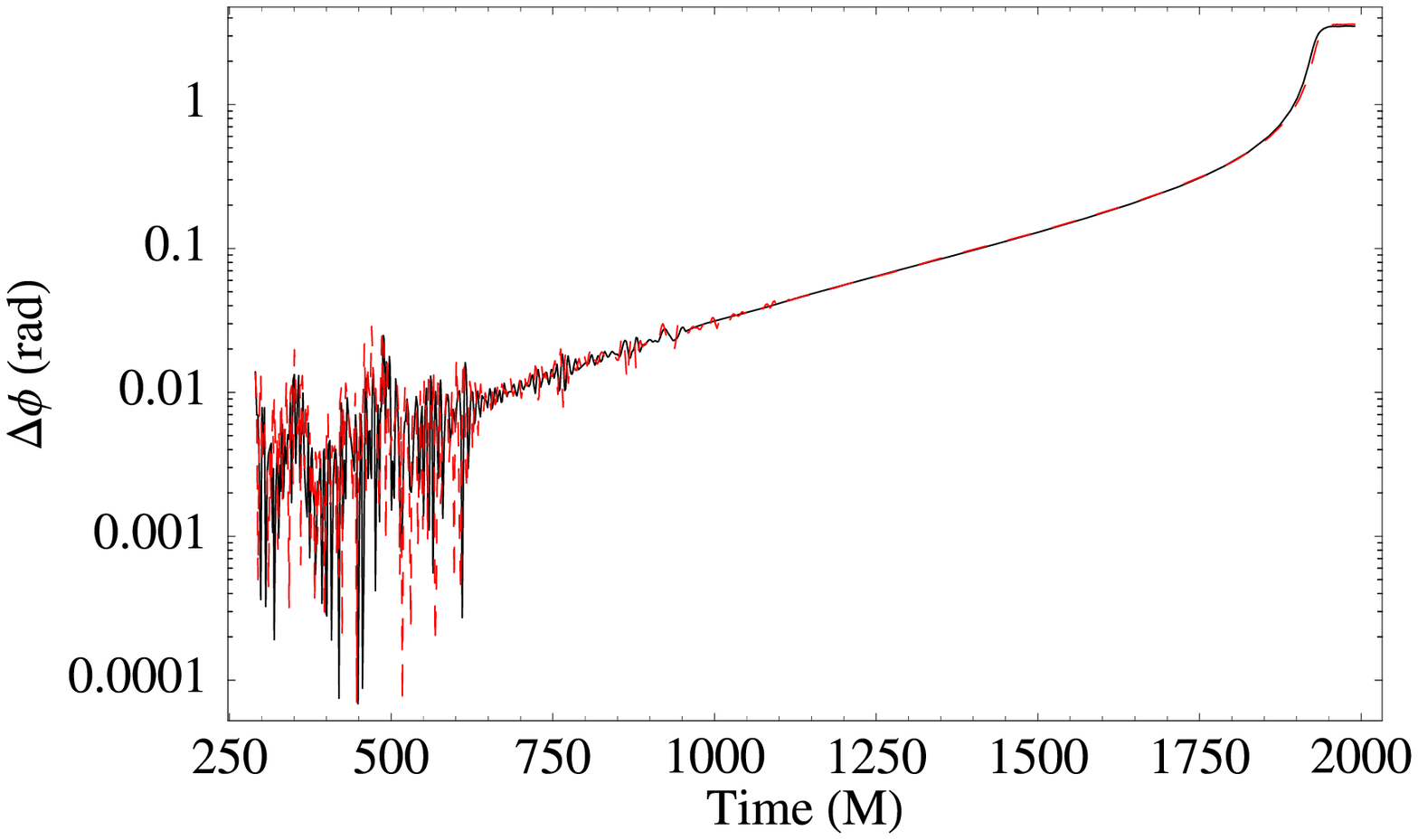}
\caption{Convergence of the phase $\phi(t)$. Differences between
simulations with $N = 64,72,80$ (see Table~\ref{tab:simulations}) are scaled
assuming sixth-order convergence. The convergence of the phase is
shown as both a standard and a logarithmic plot, to demonstrate that good
sixth-order convergence is seen throughout the simulation, except after
merger, when there is a slight drop in convergence. In the logarithmic plot
the solid and dashed lines are so close as to be almost indistinguishable.}
\label{fig:Conv1}
\end{figure}

\begin{figure}[t]
\centering
\includegraphics[height=4cm]{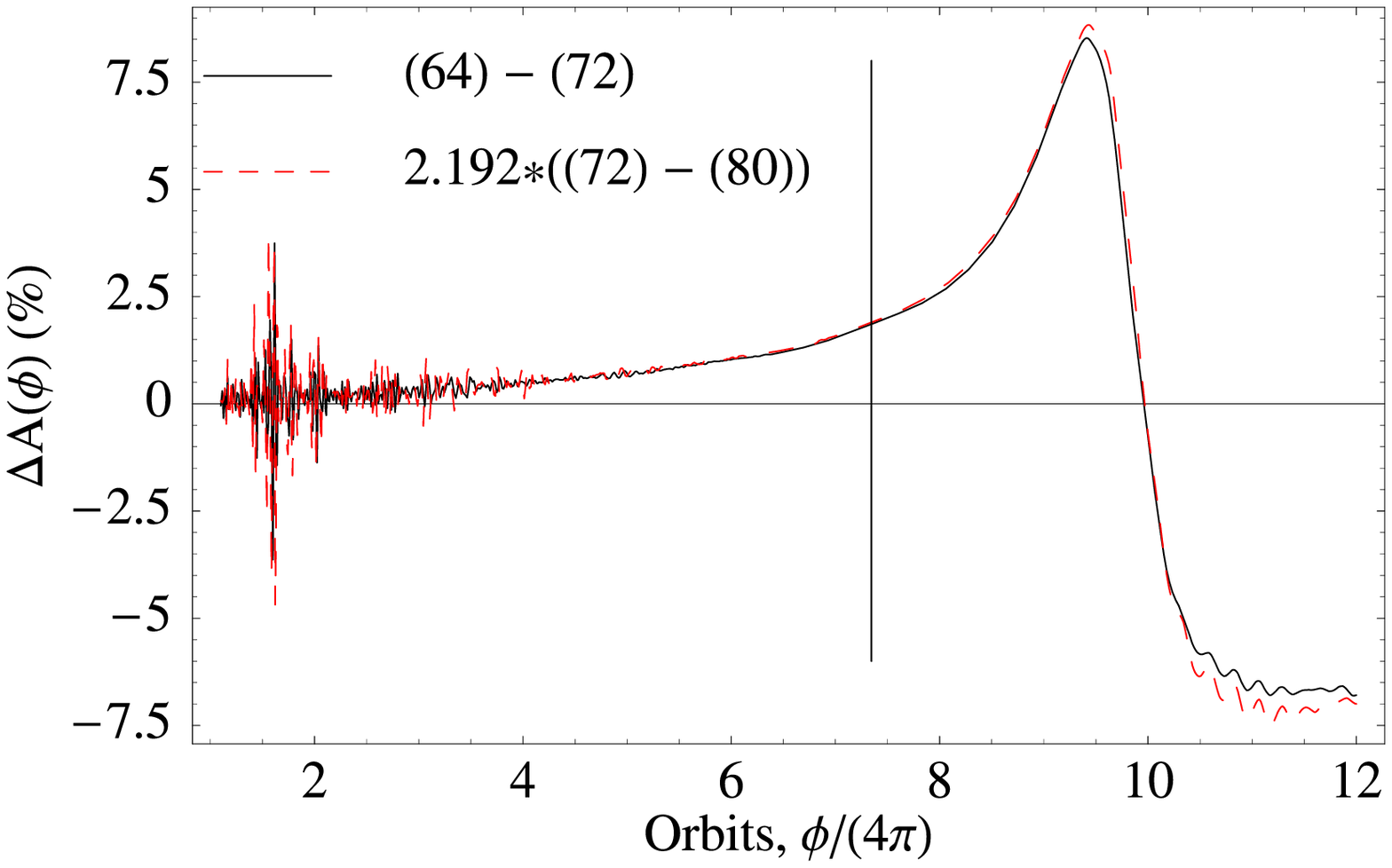}
\includegraphics[height=4cm]{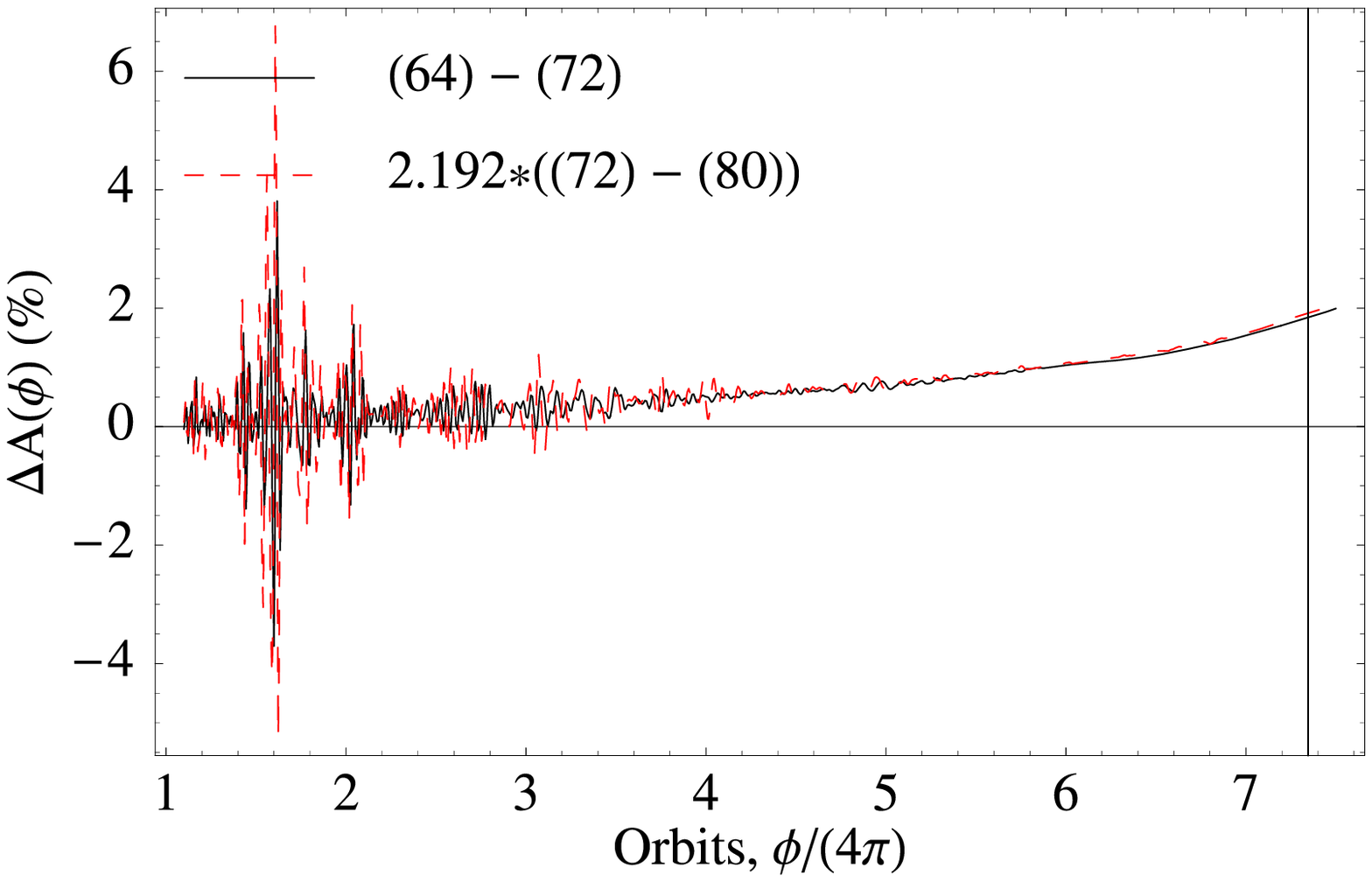}
\caption{Convergence of the amplitude $A(\phi)$. Differences between
simulations with $N = 64,72,80$ (see Table~\ref{tab:simulations}) are scaled assuming
sixth-order convergence. The $x$-axis shows $\phi/(4\pi)$, which gives a
rough estimate of the number of orbits the system has completed (at least
before merger). The phase $\phi$ is chosen to be zero at $t=0$. The
convergence of the amplitude is shown in terms of relative (percentage)
errors, to allow easier comparison with later results. A vertical line indicates 
the point at which we end our PN comparison in Section~\ref{sec:comparison}.
The lower plot zooms into the region that will be used for PN comparison.}
\label{fig:Conv2}
\end{figure}

\begin{figure}[t]
\centering
\includegraphics[height=4cm]{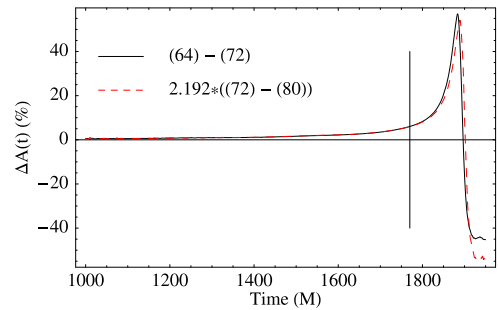}
\caption{Same as Figure~\ref{fig:Conv2}, but using $A(t)$ instead of
  $A(\phi)$. We see that the errors are far larger than for $A(\phi)$; the
  maximum error is now around 60\%, while it was only 8\% when we considered
  $A(\phi)$.}
\label{fig:Conv3}
\end{figure}

In each simulation, waves are extracted at radii $R_{ex} = 40, 50, 60, 80$ and
$90M$. Figure~\ref{fig:Conv1} shows that the phase
displays good sixth-order convergence over the course of the entire evolution. 
In order to disentangle the error in
the phase from that in the amplitude, we now consider the amplitude as a
function of {\it phase}, rather than time, $A(\phi)$, and show in
Figure~\ref{fig:Conv2} that this function is also sixth-order
convergent. For comparison Figure~\ref{fig:Conv3} also shows a convergence
plot of the amplitude as a function of time, $A(t)$, with no adjustments made
for the phase. We see that $A(t)$ is sixth-order convergent, but the errors
are almost a factor of ten larger than they are for $A(\phi)$; this
demonstrates the utility of considering $A(\phi)$ instead of $A(t)$.

The figures show the amplitude and phase from the waves extracted
at $R_{ex} = 60M$, but similar properties are seen at all five extraction
radii. Note that in these figures, and in all other relevant figures in this
paper, the 
horizontal axis displays the time from the numerical code. For example, in
Figure~\ref{fig:Conv1} the wave phase shown at $t = 1000M$ is
the phase of the wave measured at the extraction sphere at
$R_{ex} = 60M$ at code time $t = 1000M$. In subsequent plots, when some time
shifting has been applied, we indicate how this relates to the code time as
displayed in any figures.  

\begin{figure}[t]
\centering
\includegraphics[height=4cm]{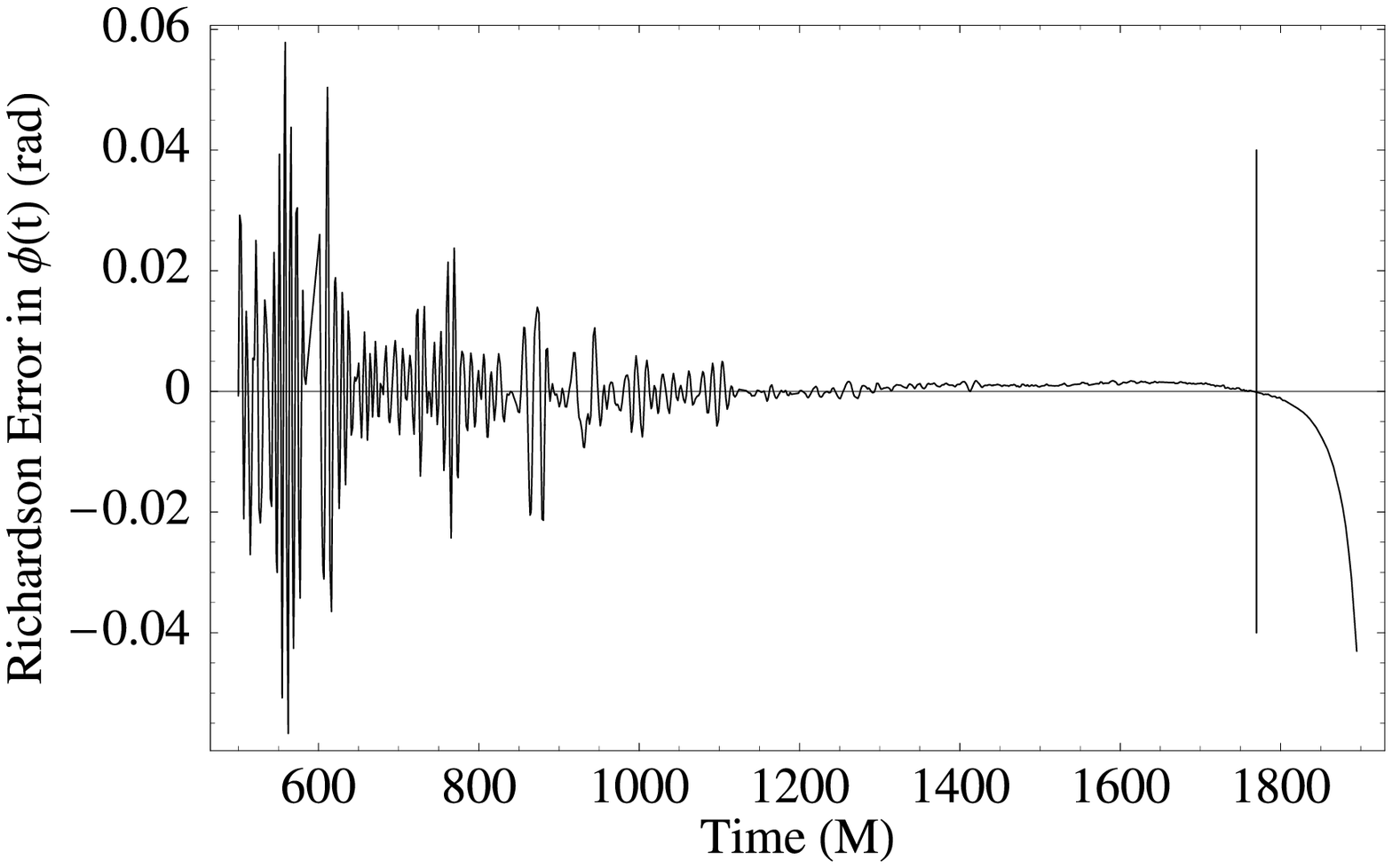}
\includegraphics[height=4cm]{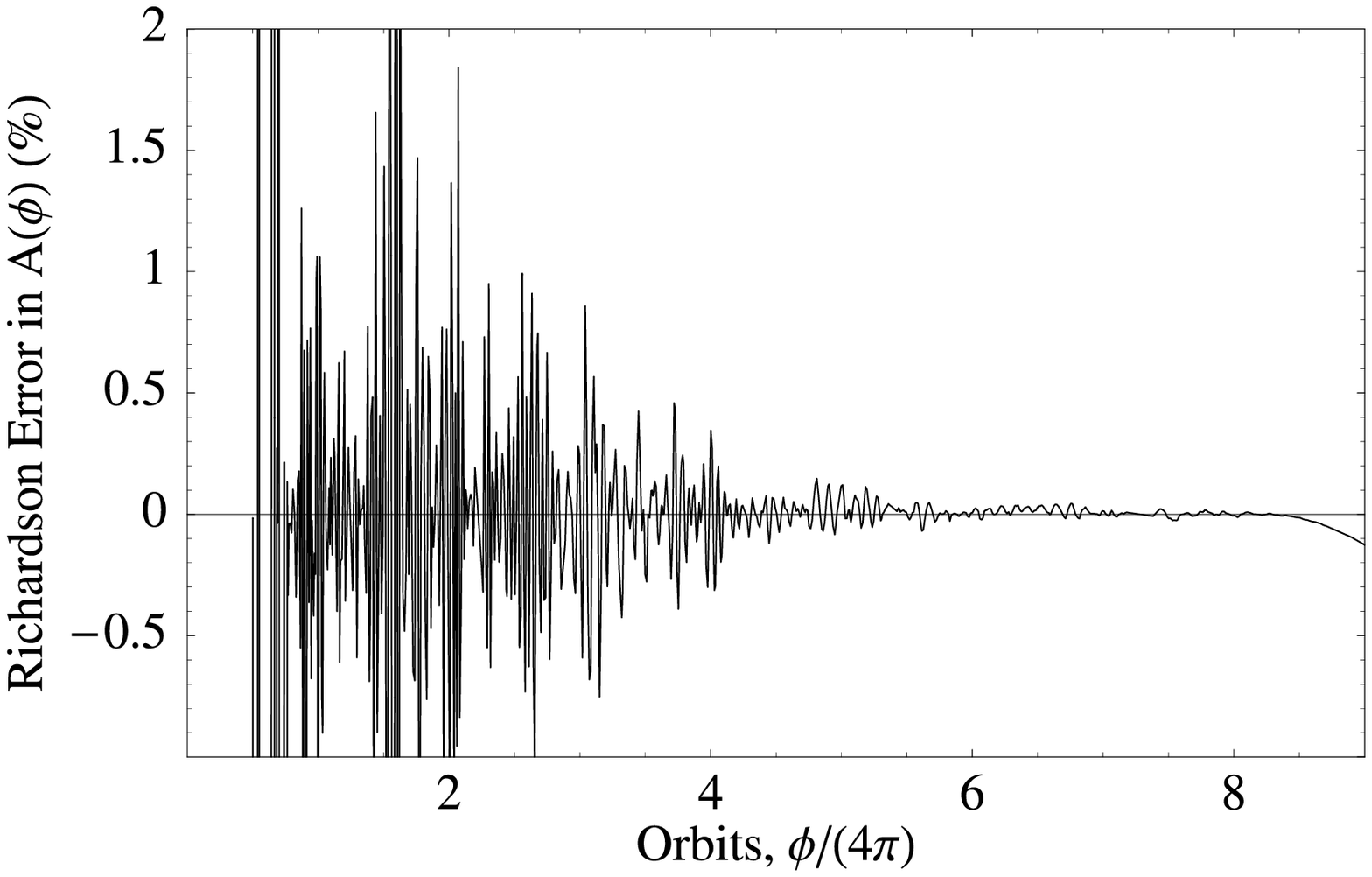}
\caption{Error in the Richardson-extrapolated functions $\phi(t)$ and
  $A(\phi)$. For the range of the simulations that will be compared with PN
  waveforms, the uncertainty in $\phi(t)$ is below 0.01 radians at most
  times, and the uncertainty in the amplitude is less than 0.5\%. At earlier
  times ($t < 500M$, which are also nominally included in the PN comparison), 
  these plots are dominated by noise and the uncertainty grows by a factor of ten.}
\label{fig:RichErrors}
\end{figure}

Given the clean sixth-order convergence of $A(\phi)$ and $\phi(t)$,
we apply Richardson extrapolation to
$A(\phi)$ and $\phi(t)$ at each extraction radius. Since we have results at three 
resolutions, we are also able to  
compute an error estimate for the Richardson-extrapolated results. If 
a function in the continuum limit is $f$, and a numerical calculation of it, $\tilde{f}$,
is sixth-order accurate, then we can write \begin{equation}
\tilde{f} = f + a_1 h^6 + a_2 h^7 + O(h^8),
\end{equation} where $h$ is the grid-spacing. With results at two resolutions, 
Richardson-extrapolation involves calculating the coefficient $a_1$ and removing
the sixth-order error to give a result that is seventh-order accurate. With results
at three resolutions, we may also calculate $a_2$, and taking the difference between
estimates of the true solution $f$ using only $a_1$ or both $a_1$ and $a_2$,
we can estimate the error in the Richardson-extrapolated result. 
These errors are shown in  Figure~\ref{fig:RichErrors} for the portion of the
simulation that will be compared with PN waveforms. We see that for $t >
500M$ the uncertainty in $\phi(t)$ is less than 0.01 radians, and the
uncertainty in $A(\phi)$ is less than 0.5 percent. At earlier times the
uncertainties grow by up to a factor of ten, due to noise in the data.

We now have amplitude and phase functions $A(\phi)$ and $\phi(t)$
for each of the 
five extraction radii, and wish to extrapolate to $R_{ex} \rightarrow
\infty$. 

\begin{figure}[t]
\centering
\includegraphics[height=4cm]{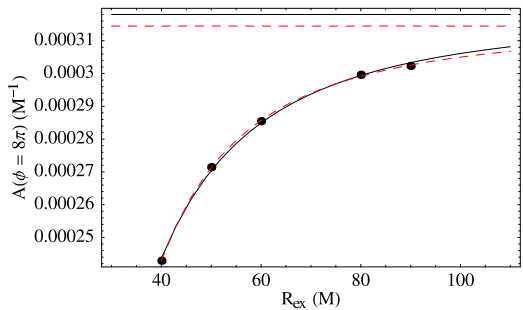}
\caption{The wave amplitude $A$ as a function of extraction radius $R_{ex}$,
  at $\phi = 8\pi$, which corresponds to $t \approx 715M$ for the wave
  extracted at $R_{ex} = 90M$. The solid line shows a curve fit of the form
  (\ref{eqn:AmpFalloff}). The dashed line shows a curve fit with an extra
  $1/R_{ex}^3$ term. The horizontal solid and dashed lines show the
  corresponding $R_{ex} \rightarrow \infty$ limits of the two curve fits; our
  uncertainty estimate in the extrapolation of the amplitude comprises the
  difference of these two values.}
\label{fig:AmpExtrap}
\end{figure}

\begin{figure}[t]
\centering
\includegraphics[height=4cm]{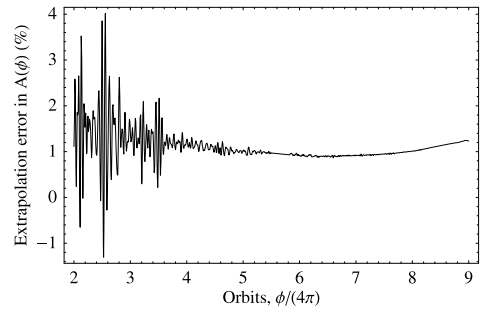}
\caption{Error in the $R_{ex} \rightarrow \infty$ extrapolated function
  $A(\phi)$. For the range of the simulations that will be
  compared with PN  waveforms, the uncertainty in the amplitude 
  is less than 2\%.}
\label{fig:ExtrapErrors}
\end{figure}

We first deal with $A(\phi)$. Since we are looking at the amplitude as a function
of phase, rather than time, the amplitudes measured at each extraction radius
are already in phase; there is no need to ``line them up'', as would be necessary 
if we looked at $A(t)$. We find that the value of the five amplitude functions is
approximated well by a quadratic function in extraction radius, i.e.,
\begin{equation}
A(\phi,R_{ex}) = A_{\infty}(\phi) + \frac{k(\phi)}{R_{ex}^2} + O\left(
  \frac{1}{R_{ex}^3} \right).
\label{eqn:AmpFalloff} 
\end{equation} In other words, the wave amplitude falls off as the square of the extraction
radius. A simple curve fit (performed at each phase $\phi$) allows us to construct
$A_{\infty}(\phi)$. Including the next fall-off term, $1/R_{ex}^3$, allows us to also 
estimate the uncertainty in the extrapolation, analogous to the method of error estimate
in the Richardson extrapolation of the discretization error. Note that although one would expect
the error to fall off as $1/R_{ex}$, our results suggest that the quadratic fall-off dominates;
this has also been observed in simulations of a particle orbiting a Kerr black hole 
\cite{Sundararajan:2007jg}. The quadratic fall-off in the amplitude error is
demonstrated in Figure~\ref{fig:AmpExtrap}.
The resulting relative error estimate as a function
of $\phi$ is shown in Figure~\ref{fig:ExtrapErrors}, and as a result of this
plot we estimate the uncertainty in $A(\phi)$ due to extrapolation to $R_{ex}
\rightarrow \infty$ as about 2\%. This dominates the
uncertainty from Richardson extrapolation ($<0.5\%$), so we also estimate the
overall uncertainty in $A(\phi)$ as about 2\%.

We now turn to the phase, $\phi(t)$. To a first approximation we expect
that the difference in the phase measured at two extraction spheres will be a 
constant.
However, the proper distance between each extraction sphere may drift due to
gauge effects. We have already seen in evolutions of the Schwarzschild
spacetime that the coordinate location of the horizon drifts depending on the
value of the $\eta$ parameter in the $\tilde{\Gamma}$-driver shift evolution
equation (see Figure 4 in \cite{Bruegmann:2006at}), and effects related to
$\eta$ have also been observed and studied in
\cite{Krishnan:2007pu,Lousto:2007db}; and it is quite 
possible that there are other gauge effects that we are not aware of. 

We have attempted to extrapolate the phase to $R_{ex} \rightarrow \infty$ by lining
up the phase at a given time, and then observing, at other times, the deviations
in the phase at different extraction radii. These deviations decrease as $R_{ex}$ 
increases, and the fall-off can be reasonably well modeled by a polynomial in
$1/R_{ex}$, and far better by a polynomial in $1/R_{ex}$ and $1/R_{ex}^2$. However,
we do not find the limit as $R_{ex} \rightarrow \infty$ to be very robust --- the results
vary depending on the choice of the time when the phases are lined up. 
(Obvious choices for this time are when the gravitational wave amplitude reaches
a maximum, near merger, or the time at which the GW frequency $M\omega$ 
equals one of the matching values that will be used in our PN comparison below.) 
As such, we do not extrapolate the phase. We instead use the phase at the largest
extraction radius, $R_{ex} = 90M$ (which we expect to be the most accurate) 
and use the phase extrapolation procedure to estimate the uncertainty in the 
phase, which we give as 0.25 radians. 

An alternative indication of the accumulated phase error of the numerical
simulations is given by the time when the amplitude of the gravitational-wave
signal reaches a maximum. This time is also seen to be sixth-order convergent,
and a similar Richardson-extrapolation error estimate as performed above gives
an uncertainty of $0.4M$ in the ``length'' of the simulation.

We are now able to construct a final waveform, \begin{eqnarray}
Re(r \Psi_{4,22}(t)) & = & A_{\infty}(\phi_{90}(t)) \cos( \phi_{90}(t) - \delta) \\
Im(r \Psi_{4,22}(t)) & = & A_{\infty}(\phi_{90}(t)) \sin( \phi_{90}(t) - \delta),
\end{eqnarray} where $\delta$ is an arbitrary phase shift, which we will apply
later when comparing with PN waveforms. The uncertainty in the wave
amplitude is about 2\%, and the accumulated phase error over the time
range we will consider is about 0.25 radians.
The time-shifting process described
earlier means that the extrapolated waveform is measured at an effective
extraction sphere with $R_{ex} = 90M$, i.e., our extrapolated waveform gives
the wave amplitude that would be measured at infinity, but at a time roughly
$90M$ after the wave was emitted from the binary system. Since we do not make
direct comparisons between quantities calculated from the gravitational
waves and quantities calculated from the puncture motion, it is not
necessary to know this ``wave travel-time'' precisely. Although not used here,
one could estimate this time using the method suggested in \cite{Fiske05}.

\begin{figure}[t]
\centering
\includegraphics[height=4cm]{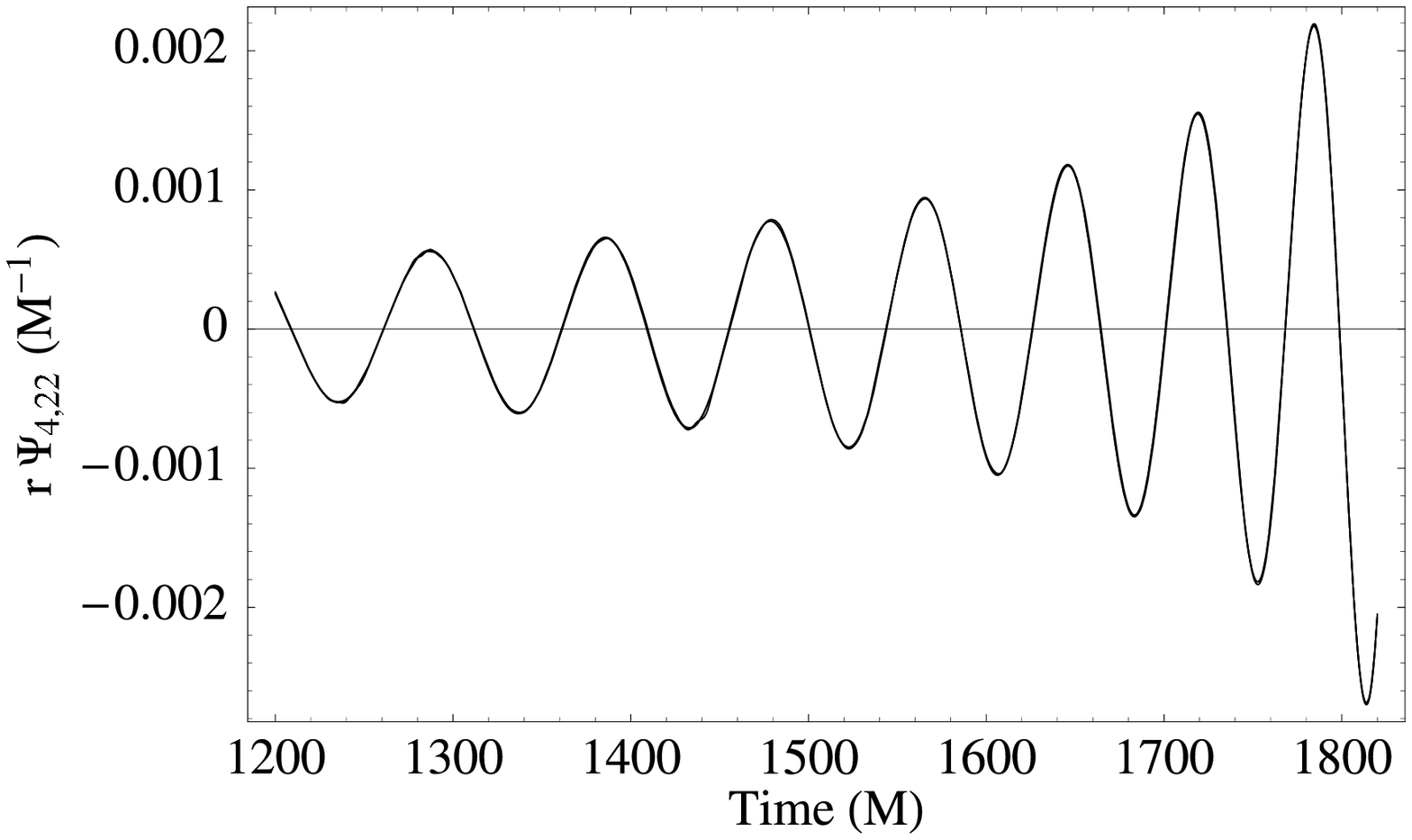}
\includegraphics[height=4cm]{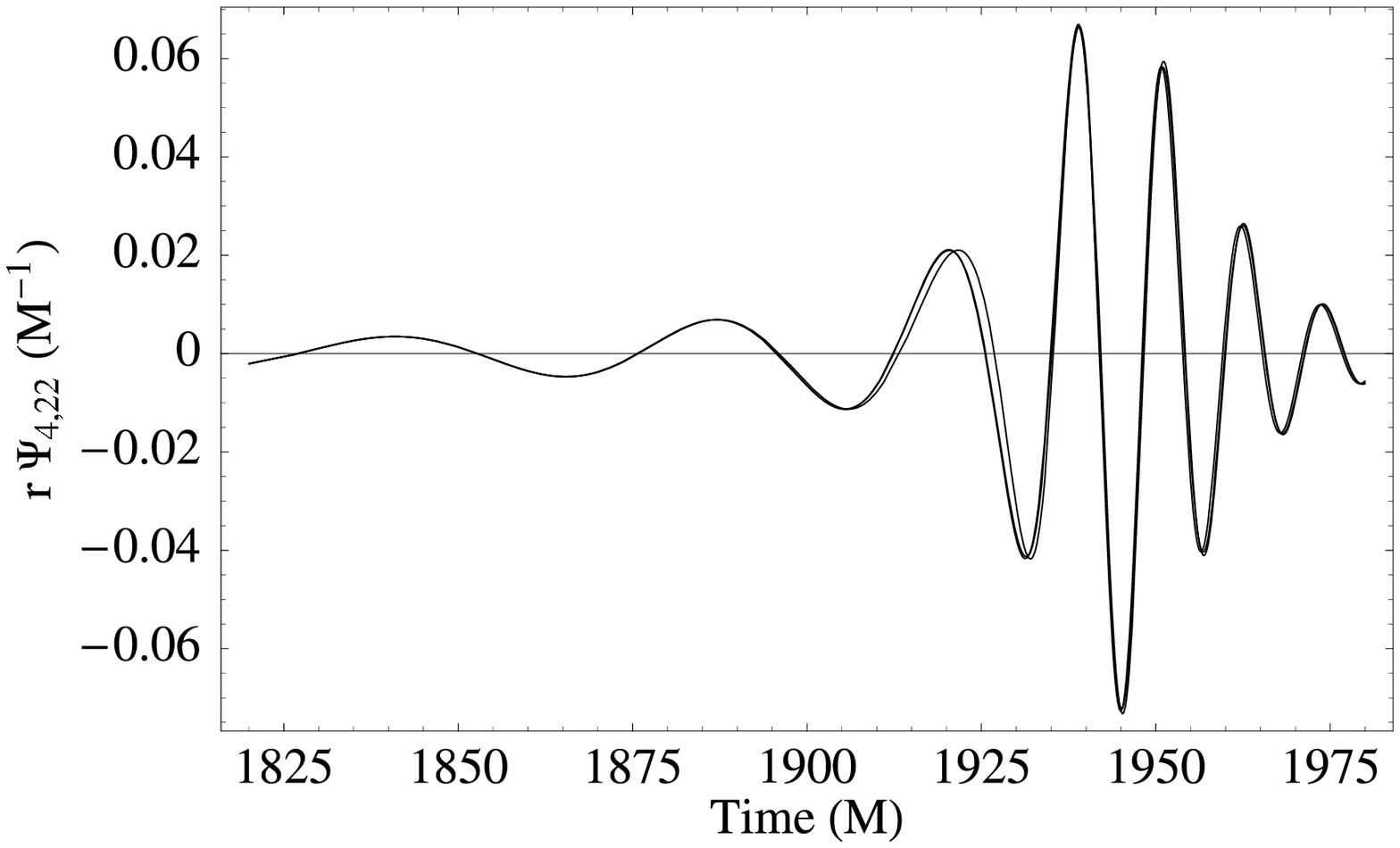}
\includegraphics[height=4cm]{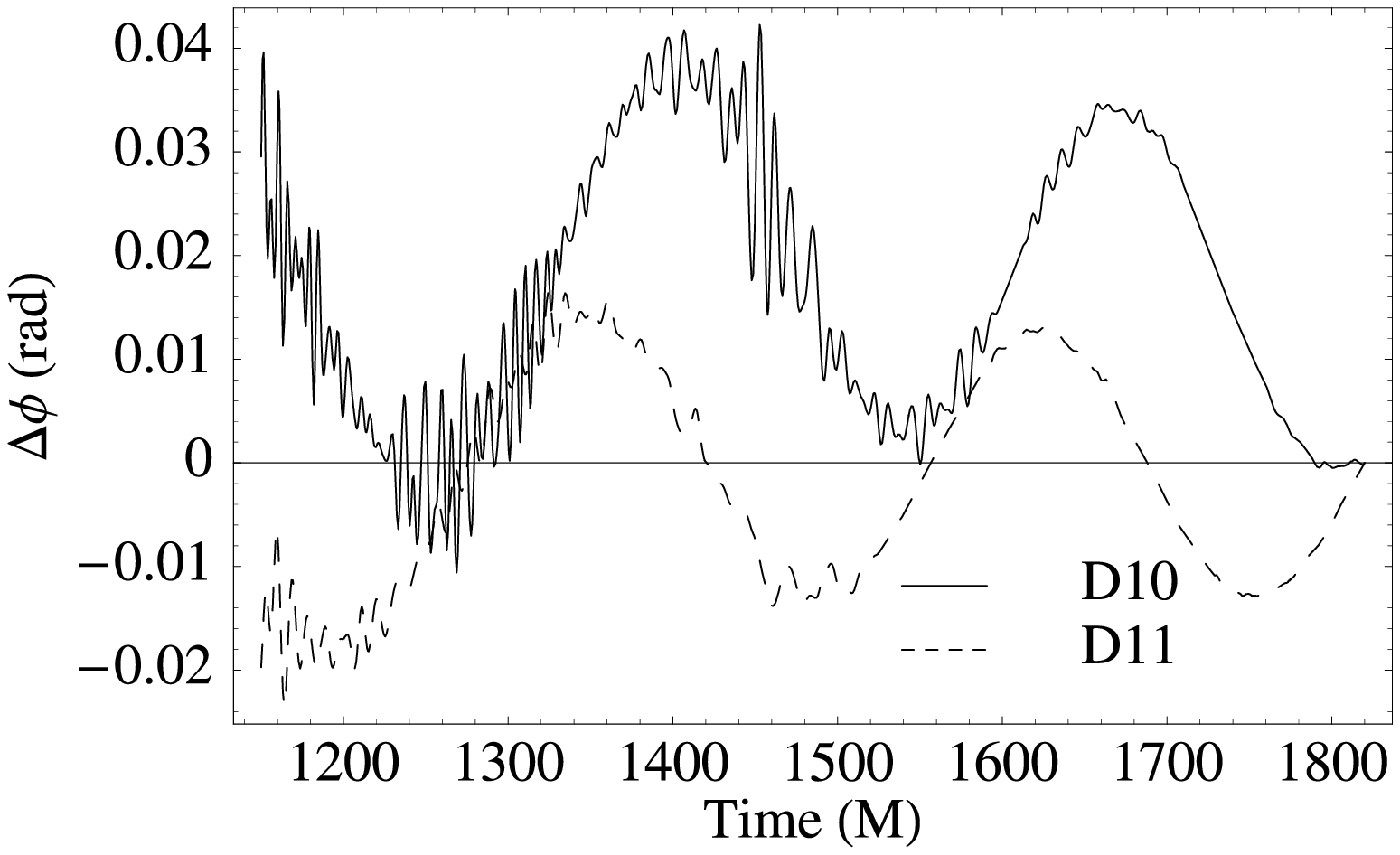}
\caption{Final waveforms from the D10, D11 and D12 runs, produced by the
  method discussed in the text and shifted in time so that their amplitude
  maxima occur at the same time. 
The three lines are not individually labeled; the main point is that their
differences are almost indistinguishable, except in the last cycle before
merger; see text. The phase disagreement with the D12 simulation is shown in
the lower panel.} 
\label{fig:RunComparison}
\end{figure}

The same procedure is applied to the simulations D10 and D11.  With a
suitable time-shift applied so that the wave amplitude maximum occurs at the
same time, the extrapolated waveforms from the three simulations are shown in
Figure~\ref{fig:RunComparison}. The waveforms lie almost perfectly on top of
each other, except in the last cycle before merger. It it at this time that we
see a ``glitch'' in the clean convergence of the phase $\phi(t)$. However, for
comparison with 3.5PN waveforms we will only be interested 
in the waveform before $t \approx 1770M$. In order to quantify the level
of agreement between the D10, D11 and D12 waveforms, we also show
in the lower panel of Figure~\ref{fig:RunComparison} the accumulated phase 
error between $t = 1200M$ and $t = 1800M$ (where $t$ is the code time of 
the D12 simulation). We see that the phase errors average to below 0.03 
radians.

\section{Comparison}
\label{sec:comparison}

\begin{figure}[t]
\centering
\includegraphics[height=4cm]{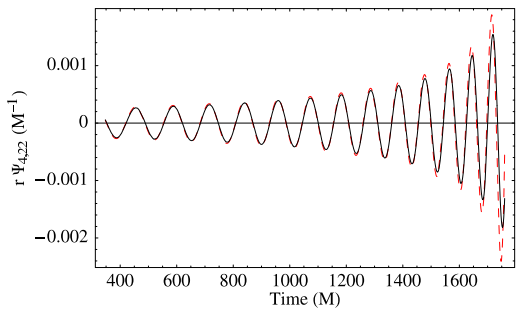}
\caption{Numerical (solid line) and TaylorT1 3.5PN (dashed line) waveforms $r
  \Psi_{4,22}$ for equal-mass inspiral. 
}
\label{fig:firstcomparison}
\end{figure}

Given the post-Newtonian (PN) and numerical-relativity (NR) waveforms discussed in
Sections \ref{sec:pn} and \ref{sec:simulations}, we are now in a position to compare 
them. We compare NR waveforms with a 3.5PN TaylorT1 waveform that was
terminated at a gravitational-wave frequency $M \omega = 0.120$, but we will only 
use it up to a cutoff frequency of $M \omega_0 = 0.1$; since the growth in
phase error in the 3.5PN waveform becomes dramatic at late times (see for example
Figure~17 in \cite{Pan:2007nw}), the smaller the choice of cutoff frequency
the better. 
Figure~\ref{fig:firstcomparison} shows the numerical D12 $r\Psi_{4,22}$ 
overlaid with the 3.5PN TaylorT1 version computed from output from the LAL
code. The figure starts at $t = 340M$, after the binary has completed one
orbit; this allows time for noise due to the junk radiation in the initial
data to leave the signal. The agreement between the PN and numerical waveforms
appears to be excellent. A similar plot (for $h_+$) is shown in Figure 1 of
\cite{Baker:2006ha}.  

The NR and PN waveforms shown in Figure~\ref{fig:firstcomparison} were ``lined up'' by
first identifying the time at which both waveforms had a given frequency $M
\omega_0$. An appropriate phase shift $\delta$ was then applied to the
numerical waveform to line up the PN and NR phases. The choice of $M \omega_0$
can have a dramatic effect on the quality of the phase agreement between the
PN and NR waveforms. Figure~\ref{fig:firstcomparison} was produced by matching 
at the beginning of the comparison region, at $M\omega = 0.0455$, which gives a far
better phase match, as we will discuss below. 

We will now discuss this subtle feature of the matching
process in more detail, before we make any conclusions about the agreement
between NR and 3.5PN TaylorT1 waveforms. 

\subsection{Phase and frequency}

The wave frequency $M\omega$ calculated from the NR waves is 
typically very noisy at early times, but becomes much smoother near
merger, when the value is higher. To allow a matching at any time in the
window of comparison, we fit a polynomial in time through the numerical
frequency to produce a smoother function. The curve fit is based on the form
of the frequency evolution in the TaylorT3 approach, i.e., a polynomial in
$(t-t_c)$, where $t_c$ is a crude estimate of the merger time (its specific
value does not strongly affect the accuracy of the fit; we used $t_c = 1927M$),
and the powers of $(t-t_c)$ that are included are
$\{-3/8,-5/8,-3/4,-7/8,-1,-9/8\}$. The use of a curve fit introduces yet
another source of error in our numerical phase, particularly at early times,
which is difficult to assess. However, the analyses below were repeated with
different fitting functions (by keeping or removing the last term in the fit,
or varying $t_c$),
and all changes in the phase results were below the stated numerical phase
uncertainty of 0.25 radians. Nonetheless, we tend to consider any matching
done at late times to be more reliable than that done at early times.

\begin{figure}[t]
\centering
\includegraphics[height=4cm]{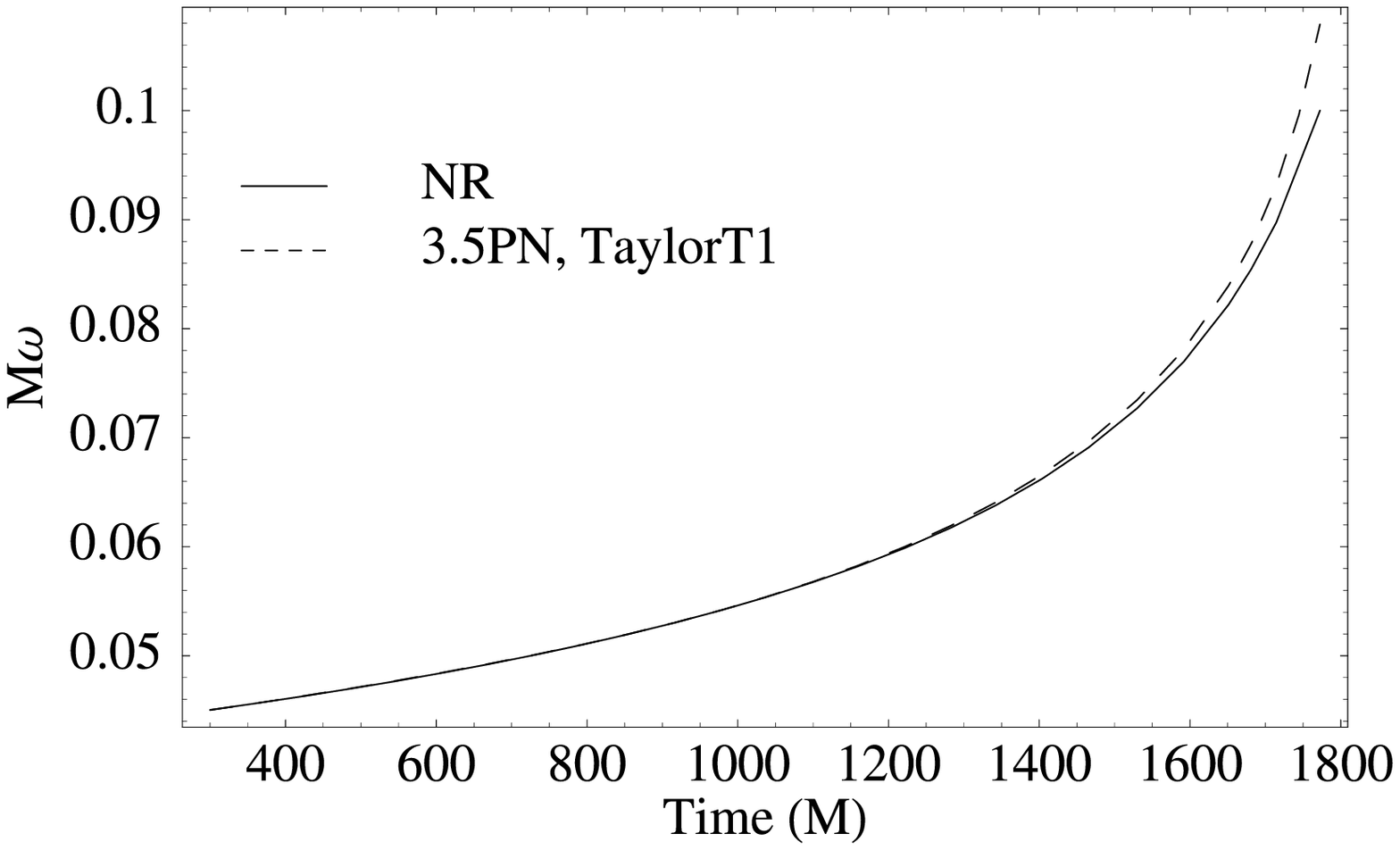}
\includegraphics[height=4cm]{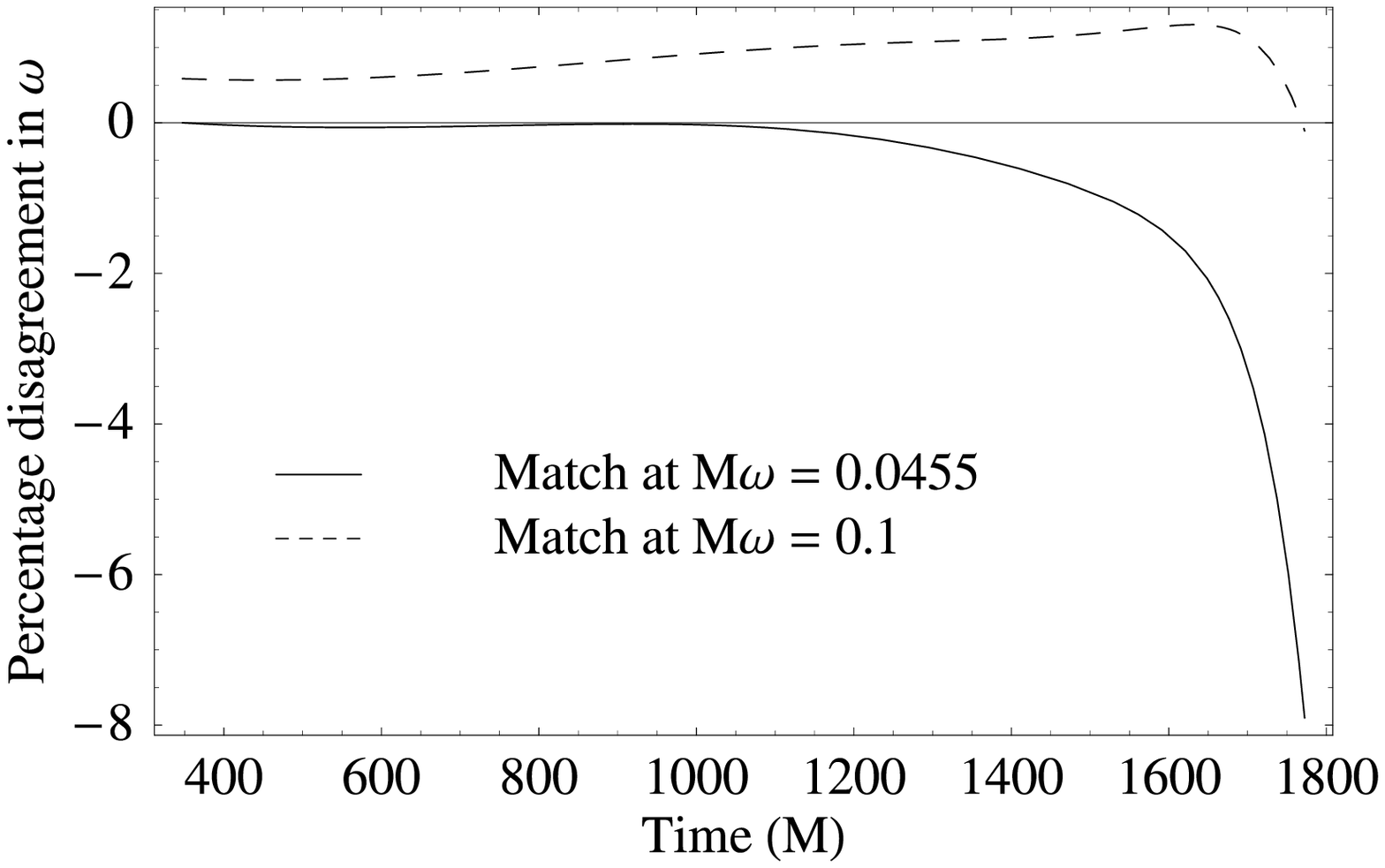}
\caption{The wave frequency $\omega$ as a function of time for the D12
  numerical-relativity and TaylorT1 3.5PN waveforms. The frequencies agree at
  $t = 347.4M$, when $M\omega = 0.0455$ The percentage disagreement between 
  the two is shown in the
  lower panel. Also shown is the frequency disagreement when matching is done
  at $M\omega = 0.1$, $t = 1772M$.}
\label{fig:OmegaComparisonEnd}
\end{figure}

On the other hand, we expect the PN phase to be most accurate at early times
--- in principle, we should be able to obtain arbitrary accuracy in the
post-Newtonian expressions by going to sufficiently early times. For that
reason we first choose to line up the frequencies at $t = 347.4M$ in code time
(recalling that this is the time when the wave reaches the extraction sphere
at $R_{ex} = 90M$), when $M\omega = 0.0455$. We are then free to make a constant phase
shift $\delta$ to align the phase of the waves; 
again aligned at $t = 347.4M$ with  $\delta = 1.367 \pi$. The agreement between
the NR and 3.5PN wave frequencies as a function of time is shown in
Figure~\ref{fig:OmegaComparisonEnd}. As can be seen in the lower panel of
Figure~\ref{fig:OmegaComparisonEnd}, the PN and NR frequencies remain close
up to around $t = 1000M$, and then drift apart and finally diverge.
Also shown (with a dashed curve) is the result of matching  at the {\it end}
of the comparison region (at 
$t = 1772M$, with $M \omega_0 = 0.1$, $\delta = 1.067 \pi$).

\begin{figure}[t]
\centering
\includegraphics[height=4cm]{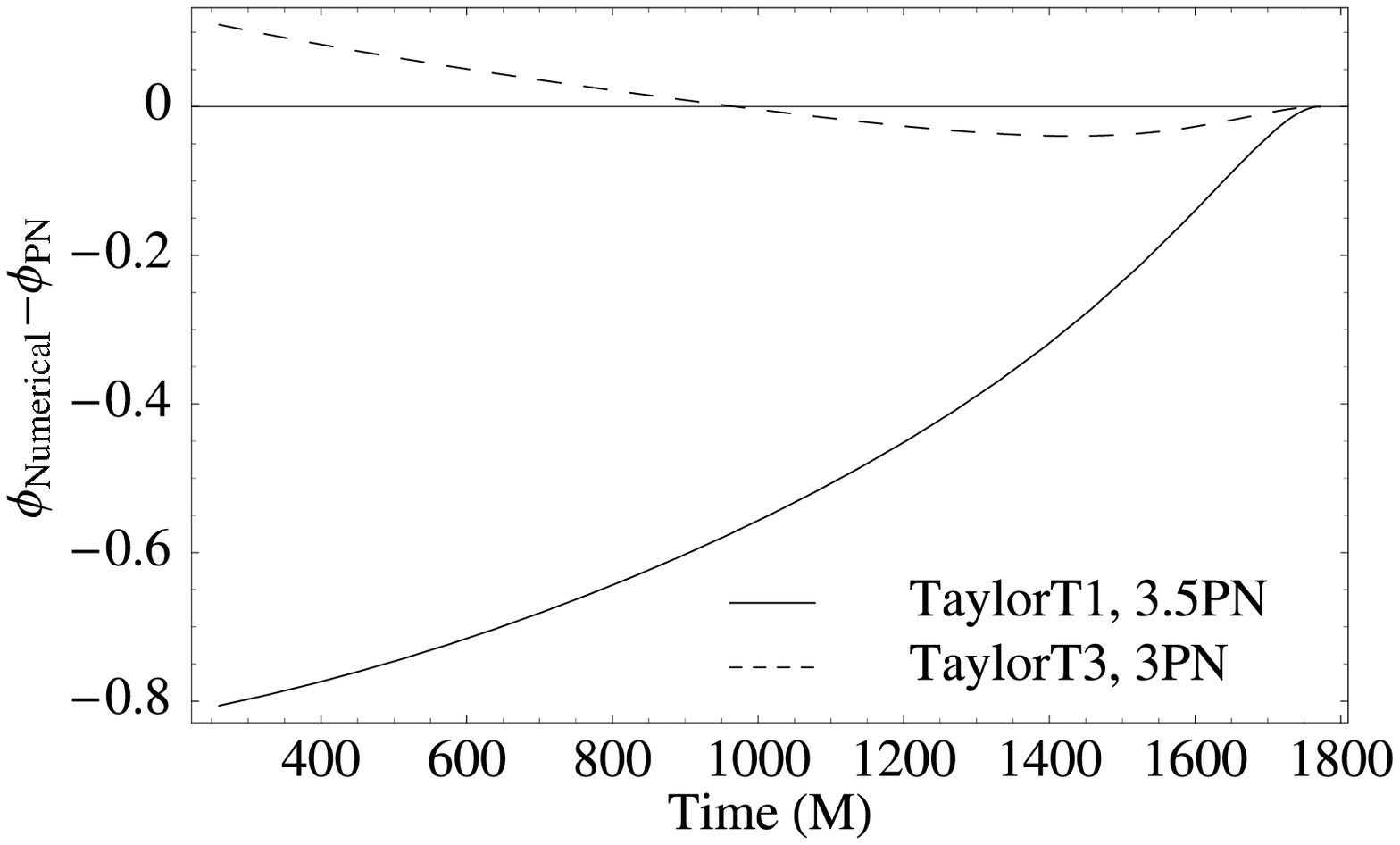}
\includegraphics[height=4cm]{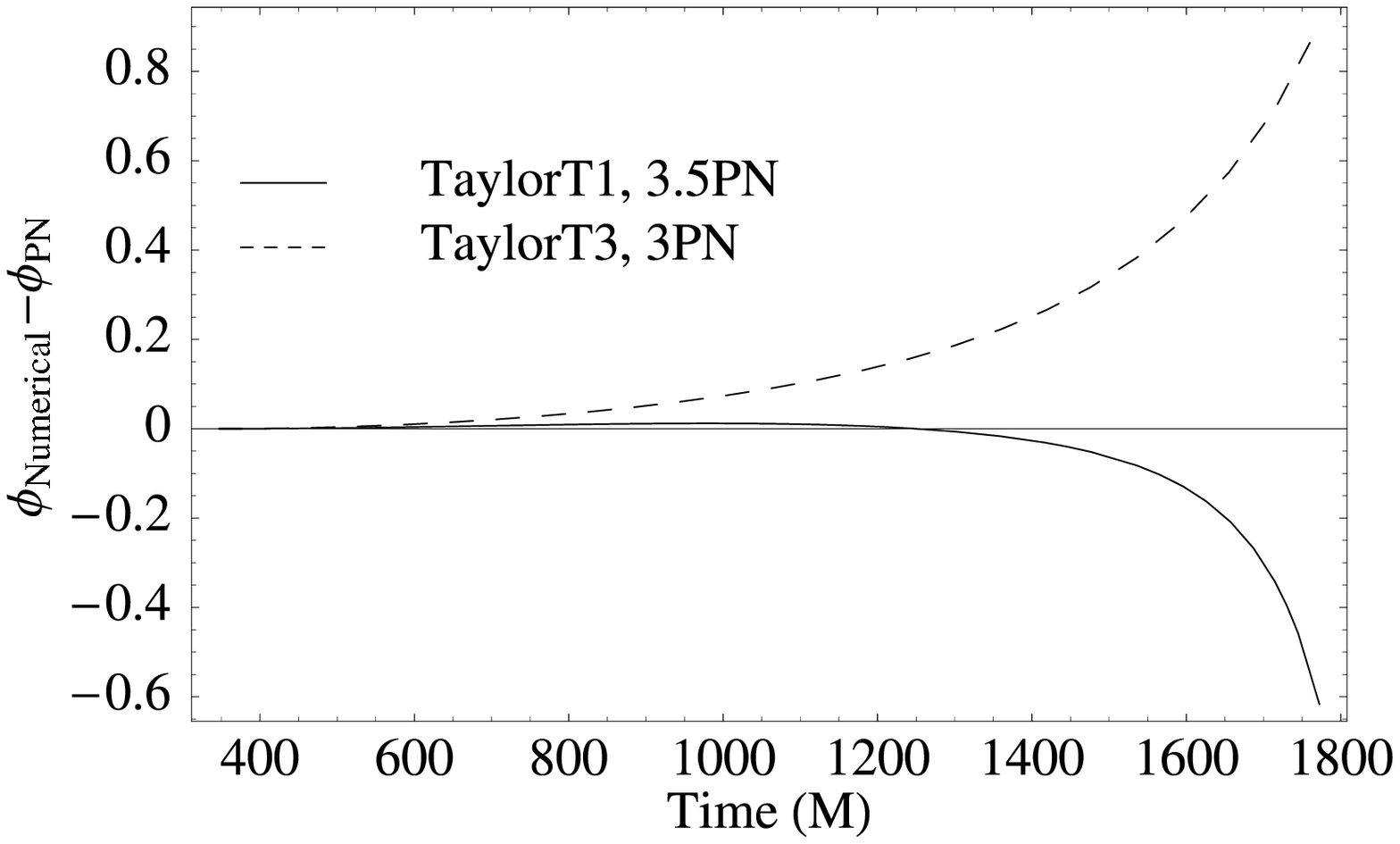}
\caption{The disagreement in the phase between NR waveforms and PN waveforms
  constructed with the TaylorT1 and TaylorT3 approximants. In the upper plot
  the phase and frequency are matched at $t = 1772M$, $M\omega = 0.1$. In
  the lower plot they  are matched at $t = 347.4M$, $M\omega = 0.0455$. We see
  that the relative merits of the two 
  approximants can appear quite different depending on the matching time.}
\label{fig:PhaseDisagreement}
\end{figure}

The corresponding results for the phase disagreement are shown in
Figure~\ref{fig:PhaseDisagreement}. Also shown is the phase disagreement
between the NR waveform and a waveform produced using the TaylorT3
approximant. In order to line up the phase and frequency of the T3 waveform,
we choose an appropriate coalescence time $t_c$ and phase constant
$\phi_0$.  

Figure~\ref{fig:PhaseDisagreement} demonstrates that the different choices of
matching frequency can give entirely different impressions of the relative
merits of the T1 and T3 approximants: when the waves are matched at $t =
1772M$, the accumulated phase disagreement between the T3 approximant and
numerical results is about 0.1 radians. When the matching is done at $t =
347.4M$, the accumulated T3/NR phase disagreement is almost 1 radian. In both
cases the T1/NR disagreement is comparable, although this is purely an
accident of the matching frequencies that were chosen. It should be clear from
the lower panel of Figure~\ref{fig:PhaseDisagreement} that if we cut off the
comparison at $t = 1000M$, the T1/NR accumulated phase error will be very
small. Similarly, for matching purposes, one could optimise the matching time
to give the smallest phase error --- for the T1 waveforms, we can for example
match at $M\omega \approx 0.075$ and achieve a phase agreement within
numerical uncertainty. 

We repeat that for the purposes of comparing PN and NR phases, the match at
$M\omega = 0.1$, when the numerical data is relatively free of noise, is the most
trustworthy. The matches at earlier times are less accurate and mainly serve
to illustrate the general trend in the disagreement between PN and NR phases:
the frequency disagreement changes sign 
(as shown in Figure~\ref{fig:OmegaComparisonEnd}), and, depending on the
approximant used and the chosen matching time and frequency, the phase
disagreement may behave as in the T3/NR curve in
the top panel of Figure~\ref{fig:PhaseDisagreement} or the T1/NR curve in in
the lower panel, and exhibit a local maximum, which allows us to optimize the
phase disagreement.

\begin{figure}[t]
\centering
\includegraphics[height=4cm]{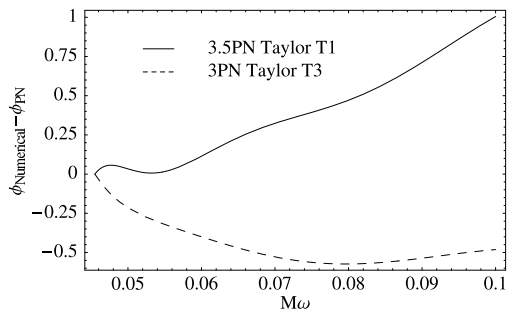}
\caption{The disagreement in the phase between NR waveforms and PN waveforms
  constructed with the TaylorT1 and TaylorT3 approximants, but now shown as a
  function of GW frequency $M\omega$.}
\label{fig:PhaseVsOm}
\end{figure}

We may produce yet another picture of how T1 and T3 behave by plotting the
phase disagreement versus the wave frequency $M\omega$, as done in  \cite{Baker:2006ha}.
This is shown in Figure~\ref{fig:PhaseVsOm}, which now suggests that T3 behaves
far better than T1. 

What are we to conclude, then, about the phase agreement between NR and T1 or
T3 PN waveforms? Due to a turning point in the evolution of the frequency
diagreement, we are left with a great deal of freedom about how to match the
frequency and phase. We find, in the frequency range that we consider, that
the minimum accumulateed phase disagreement that we can achieve is about 0.2
(or 0.15) radians using either the T1 (or T3) approximants (see
Figure~\ref{fig:PhaseDisagreement}). By contrast, the maximum disagreement
between the NR and PN phases over the comparison region is about 1 radian,
although since this results from a matching at early times, and the phase
disagreement is diverging at the end of the comparison region, this value has
a large uncertainty.

In a matching between NR and PN waveforms (as performed in, for example,
\cite{Ajith:2007qp,Ajith:2007kx}), we naturally choose to match in such a way
that the phase disagreement 
is minimized. We could easily have found that the PN phase evolution disagreed
so badly with the NR phase evolution that it was not possible to achieve an
accumulated phase disagreement of less than, for example, 1 radian. However, 
the minimum accumulated phase disagreement that we can achieve is about 0.2
radians, which is also within the phase uncertainty of the numerical
waveforms. 

We therefore
conclude that we can match the phase within the numerical uncertainty over
the frequency range we have considered ($M\omega = 0.0455$ up to $M\omega =
0.1$), and that the accumulated PN and NR phase disagreement has an upper
bound of roughly 1 radian. We expect that matching at even earlier times (using longer
simulations) would make the matching clearer, although this will also require
more accurate simulations and larger radiation extraction radii to resolve the
lower-frequency, lower amplitude waves.

\subsection{Amplitude}

We now turn to the amplitude. 

\begin{figure}[t]
\centering
\includegraphics[height=4cm]{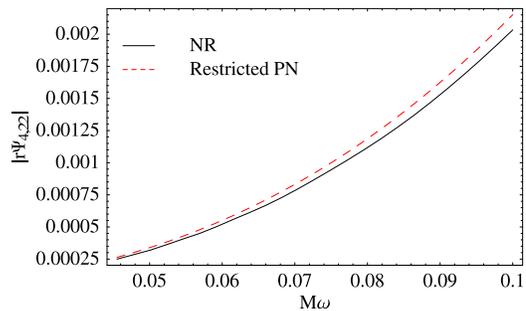}
\caption{NR and restricted PN amplitudes of $r\Psi_{4,22}$.
}
\label{fig:AmplitudeNR3PN}
\end{figure}

\begin{figure}[t]
\centering
\includegraphics[height=4cm]{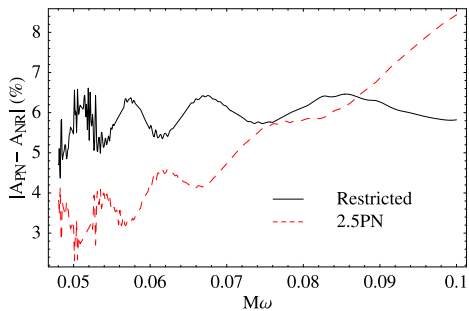}
\caption{Percentage disagreement between the NR wave amplitude and that of
the PN waves with the restricted (quadrupole) and 2.5PN amplitude
treatments. At low frequencies the 2.5PN amplitude agrees with the NR
amplitude within numerical uncertainty.}
\label{fig:AmpErrorNR3PN}
\end{figure}

Figure~\ref{fig:AmplitudeNR3PN} shows the amplitude of $r\Psi_{4,22}$ from
NR and restricted PN waves, plotted as a function of GW frequency
$M\omega$, so that the choice of PN approximant does not affect the result. The
amplitude of the restricted 3.5PN wave is larger than that for the NR 
wave. Figure~\ref{fig:AmpErrorNR3PN} shows the percentage disagreement between the 
 restricted PN and NR wave amplitudes over the same frequency range. 
The disagreement is of the order of 6\%. Since the uncertainty in the NR wave
amplitude is below 2\%, at least for $M\omega > 0.05$, we cannot
ascribe this disagreement entirely to 
numerical error. If we assume that the NR wave more closely models the correct
physics of the binary system, then the restricted PN (quadrupole) amplitude
over-estimates the  amplitude by between 4 and 8\% in this frequency range. 

So far we have compared our NR waveforms with {\it restricted} 3.5PN waveforms,
meaning that the amplitude in the gravitational-wave strain is proportional to 
$x = (M \omega/2)^{2/3}$. (The factor of two signifies that $x$ deals with the 
frequency of the black holes' motion, not the frequency of the waves; the two 
frequencies are assumed to be related by a factor of two.) If we move beyond
restricted waveforms, and model the 
amplitude up to 2.5PN order (i.e., with terms up to $x^{7/2}$)
\cite{Arun04,Baker:2006ha}, we find greater disagreement at higher frequencies,
but at low frequencies the 2.5PN amplitude shows better agreement with the 
NR amplitude. The 2.5PN amplitude disagreement at $M\omega = 0.0455$ is
between 1\% and 5\%; the PN and NR amplitudes now agree within numerical
uncertainty. This is also shown in Figure~\ref{fig:AmpErrorNR3PN}.  

As we have said, the amplitude disagreement between the NR and restricted PN
amplitudes is roughly constant over the frequency range $M\omega = 0.05$ to
$M\omega = 0.1$. This suggests that if we are content with these levels of
error when matching numerical and PN waveforms, the large 
number of cycles in the D12 simulation is not necessary. A combined PN-NR waveform
could be produced by applying a scale factor, as is done using different
approaches in \cite{Ajith:2007qp,Ajith:2007kx} and \cite{Pan:2007nw}, and clearly only a few
cycles shared by the NR and 3.5PN waveforms are needed to determine the scale
factor. We may now ask: can we get away with a numerical simulation that
starts at, for example, $D = 9M$, and yields a waveform that (neglecting the
first orbit) shares four cycles with the 3.5PN wave? 

\begin{figure}[t]
\centering
\includegraphics[height=4cm]{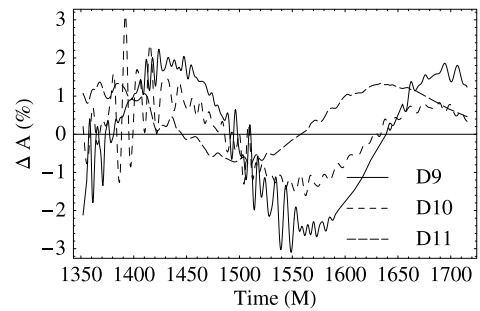}
\caption{The percentage difference in amplitude between the D12 simulation and
  the D9, D10 and D11 simulations. The D9, D10 and D11 amplitudes show small
  oscillations around the D12 value, but recall that the disagreement between the D12 
  and restricted 3.5PN amplitudes was 6\%.
}
\label{fig:AmpComparison}
\end{figure}

Figure~\ref{fig:AmpComparison} shows the relative disagreement in amplitude
between the D12 simulation and the D9, D10 and D11 simulations. 
There are small oscillations around the D12 values, but these are smaller
than the average amplitude disagreement between the NR and restricted 3.5PN wave
amplitudes, and we expect that it will be possible to calculate a 
suitable scale factor for matching the NR and 3.5PN waves. 
We conclude then that simulations starting as close as $D = 9M$ and simulating
about 4.5 orbits should be enough to match to
restricted 3.5PN waveforms for many applications. To make this clearer: any 
GW data analysis application that requires an amplitude accuracy of at most 5\% up to
the last four cycles, and an amplitude accuracy of better than 2\% from that
point through merger and ringdown, will require only short (4.5 orbit)
numerical simulations to match to PN waveforms. 

This result is attractive from a computational point of view. The D9 
simulation ran in 750 CPU hours (two and a half days of wall clock time on 12  
processors), while the highest resolution D12 simulation required
10,500 CPU hours (18 days on 24 processors).  When producing many waveforms
for use in gravitational-wave data analysis, we would much rather only have
to perform the two-and-a-half-day simulations.

Of course, in the case of equal-mass binary inspiral, we have already presented
waveforms that cover far more than four cycles before merger. The important
question is whether similarly short simulations will be adequate beyond the
equal-mass nonspinning case, and that will be the subject of future work.

\subsection{Comparison with eccentric waveforms}
\label{sec:eccentricity}

The numerical simulations discussed in the previous sections modeled 
equal-mass inspiral with negligible eccentricity, starting from the initial
parameters introduced in \cite{Husa:2007ec}. The eccentricity of the D12 
simulation is estimated as $e < 0.0016$. In contrast, one could use standard
``quasi-circular orbit'' parameters (i.e., parameters calculated with the 
assumption that $\dot{r} = 0$), which lead to inspiral with a small but 
noticeable eccentricity. We now consider a set of simulations with the same
parameters as the D12 runs, but using initial parameters calculated using
the 2PN-accurate expression used in \cite{Bruegmann:2007a} (and based
on the results in \cite{Kidder1995}); we denote this simulation ``QC12''. 
We apply the same extrapolation procedure 
as described in Section~\ref{sec:simulations} to produce the final waveform
that we analyze. 

\begin{figure}[t]
\centering
\includegraphics[height=4cm]{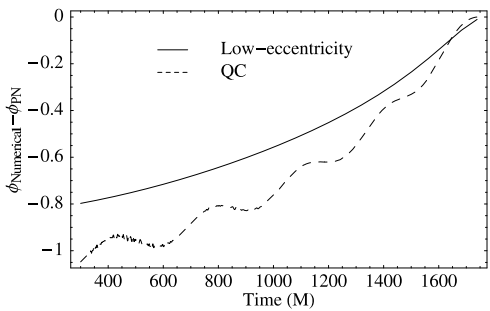}
\caption{The same quantities as in Figure~\ref{fig:OmegaComparisonEnd}, this time
comparing both the D12 and QC12 wave phase with that of the TaylorT1 3.5PN 
waves. The disagreement between the 3.5PN and QC phases displays
clear oscillations, presumably due to the eccentricity in the QC12 simulation.
}
\label{fig:PhaseErrorQC}
\end{figure}

\begin{figure}[t]
\centering
\includegraphics[height=4cm]{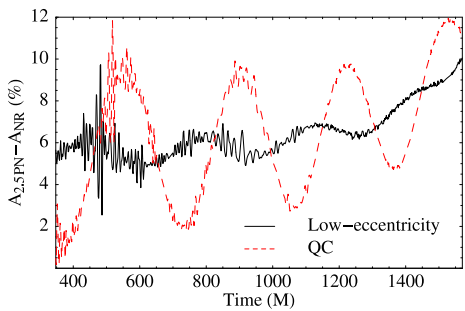}
\caption{Percentage disagreement between restricted 3.5PN and NR wave amplitudes,
for both D12 and QC12 simulations. The disagreement between the QC12
and 3.5PN wave amplitudes is clearly dominated by eccentricity. The 
low-eccentricity D12 simulation is necessary to identify the 
error in the restricted 3.5PN wave amplitude.
}
\label{fig:AmpErrorNR3PNQC}
\end{figure}

Figure~\ref{fig:PhaseErrorQC} shows the same comparison with the TaylorT1
3.5PN wave phase as in the upper panel of Figure~\ref{fig:PhaseDisagreement},
but now displaying results from both the D12 and QC12 simulations. The
accumulated phase disagreement for the QC12 simulation is larger. The
disagreement with the 3.5PN phase 
also shows oscillations that are presumably due to eccentricity. 
A similar effect can be
seen in Figure~\ref{fig:AmpErrorNR3PNQC}, which shows the percentage 
disagreement in wave amplitude. The amplitudes are now shown as functions of
time; if we use $M\omega$ as before, the eccentricity effects are not
visible. The low-eccentricity D12 waveform has been matched with the PN
waveform at $M\omega = 0.0455$, and the QC12 waveform is matched with the D12
waveform so that their amplitude maxima occur at the same time. The amplitude
disagreement between the D12 simulation and the 2.5PN amplitude is slightly
different to that shown in Figure~\ref{fig:AmpComparison}; this is due to
parameterizing the amplitude with time instead of frequency --- the PN/NR
frequency disagreement means that there is not a 1-1 relationship between the
two plots. However, the results are consistent within the 2\% uncertainty in
the numerical waveform amplitude. 

The disagreement in amplitude between the
3.5PN and QC12 results oscillates between 2\% and 10\% at early times. From the QC12 
simulation alone, we may guess that the error in the 3.5PN wave amplitude is the 
average of  this curve, i.e., around 6\%, but may also guess that the disagreement
might go away if the eccentricity were removed. The D12 simulation, which displays far 
less eccentricity, confirms the first guess: there is strong numerical
evidence that the restricted 3.5PN wave amplitude really does
 disagree with fully general-relativistic results by about 6\%.

\subsection{Comparison of the black hole coordinate motion}
\label{sec:tracks}

To initialize our numerical simulations, we have set the initial momenta
of the black holes to values we have obtained from a post-Newtonian inspiral
calculation as described in \cite{Husa:2007ec}. The inspiral calculation
starts at an initial separation of $D = 40M$ with momenta given by the 3PN-accurate
quasicircular-orbit formula given in \cite{Bruegmann:2006at}. When the
inspiral reaches the separation $D = 12M$, the momenta are read off from the
solution and given the values shown in Table~\ref{tab:parameters}. 

In this section we compare a full GR simulation that uses those parameters with
continuing the PN inspiral from $D=12M$.

\begin{figure}[t]
\centering
\includegraphics[height=4cm]{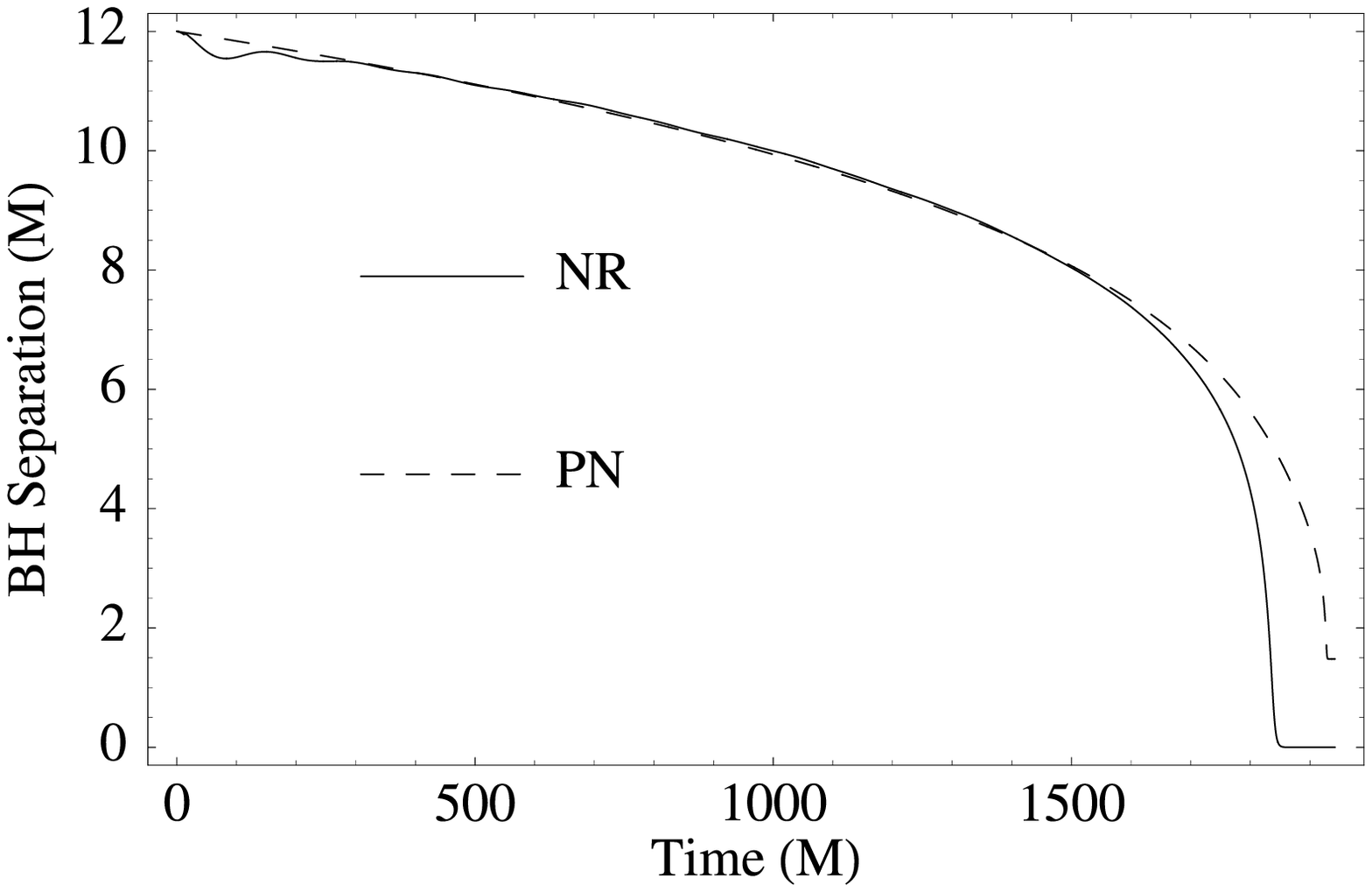}
\includegraphics[height=4cm]{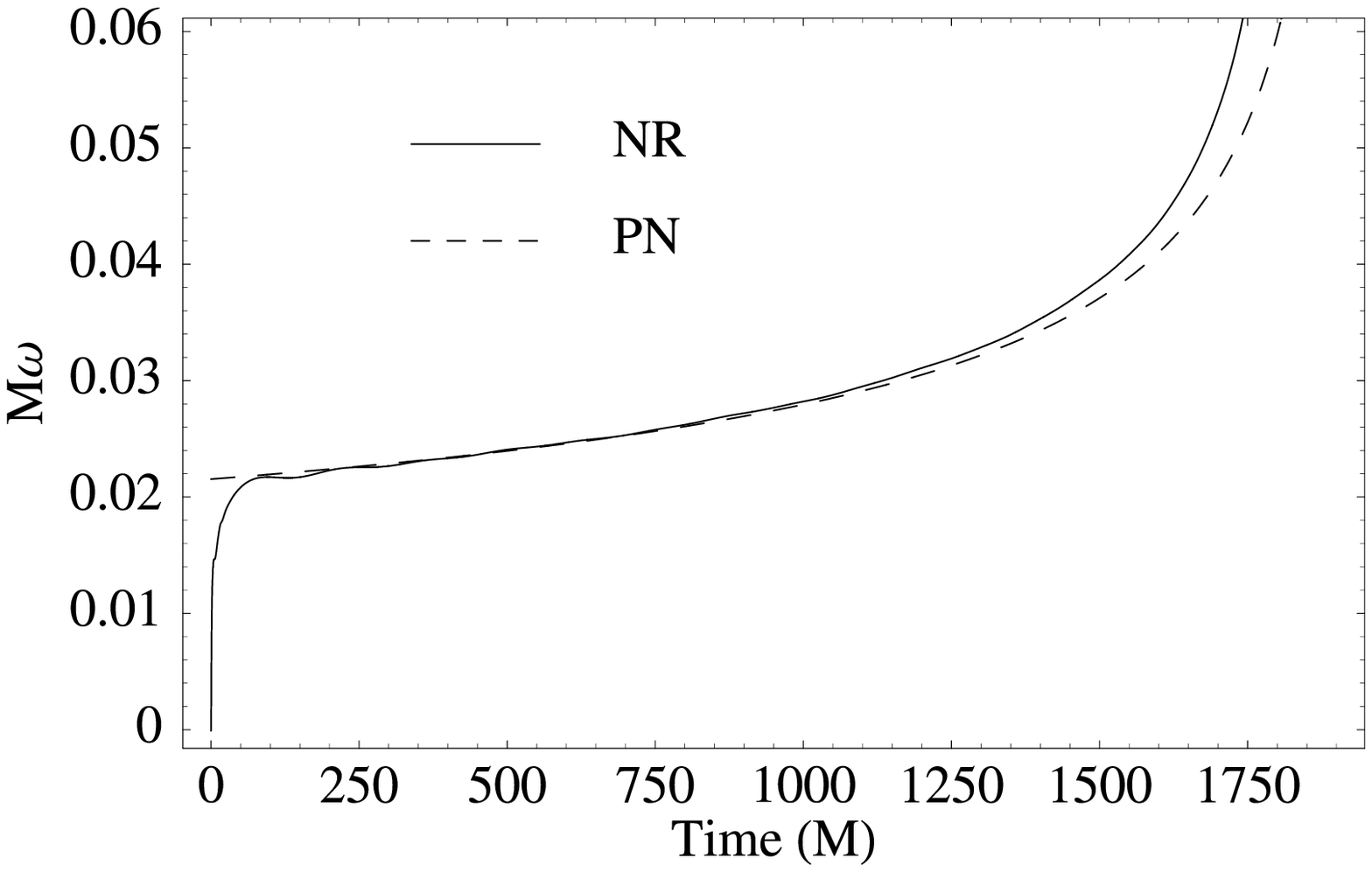}
\caption{Orbital coordinate motion of the D12 numerical relativity evolution compared with
	a PN evolution with the same initial parameters. In both panels the PN evolution
        is drawn as a dashed line. Top panel: the separation of the black holes
(the puncture position in the full NR case).
Bottom panel: the coordinate angular velocity.}
\label{fig:tracks}
\end{figure}

\begin{figure}[t]
\centering
\includegraphics[height=4cm]{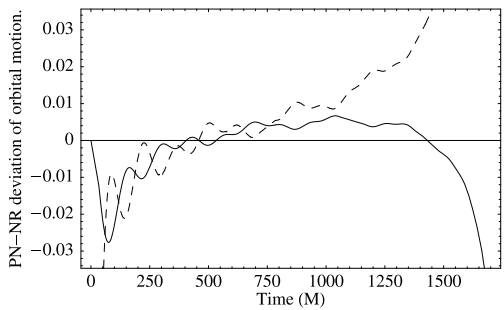}
\caption{Orbital coordinate motion of the D12 numerical relativity evolution compared with
        a PN evolution with the same initial parameters. Dashed line:
        $(\omega_{NR} - \omega_{PN})/\omega_{PN}$, 
full line: $(D_{NR} - D_{PN})/D_{PN}$ 
}
\label{fig:tracks_diff}
\end{figure}

The coordinate separation of the
black-hole punctures was chosen as the coordinate separation of the post-Newtonian
inspiral, which we have computed in ADMTT-coordinates.
This is motivated by the fact that the PN solution in the ADMTT gauge for a
two-body system agrees with our Bowen-York puncture initial data up to 2PN
order (see, for example, the explicit solutions in Appendix A of
\cite{Jaranowski98a}). It is therefore interesting to know when the use of the
ADMTT gauge breaks down in our evolutions. An indirect check is
straightforward: we compare the  PN and full NR puncture separation,
as seen in Figures~\ref{fig:tracks} and \ref{fig:tracks_diff}. 
Using the  D12 simulation, we find that both the separation and orbital phase agree 
very well from $D = 11M$ up to $D = 8M$, or from $t = 300M$ (the time to complete
the first orbit) to $t = 1500M$. Put another way, the PN and full NR
coordinate separation agrees until about 3 orbits before merger. 


\section{Discussion}

We have simulated nine orbits, merger and ringdown of an equal-mass binary, 
and extracted waveforms of sufficient accuracy to make a detailed comparison
with post-Newtonian (PN) waveforms. The uncertainties in the numerical waveforms are
dominated by the close extraction radii, and not finite-difference errors. The
PN waveforms that we focused on 
were those generated by the TaylorT1 3.5PN procedure implemented in the
LSC Applications Library (LAL), which is a candidate for use in gravitational-wave
searches in detector data; we also compared with the TaylorT3 approximant. 
We find that the phase of the TaylorT1 3.5PN waveform can be matched to agree with
the numerical phase to within numerical uncertainties, and the upper bound of
the accumulated phase disagreement is on the order of 1 radian. The restricted PN amplitude
overestimates the numerical 
value $(6\pm2)$\%. We have found that
the ratio of the restricted PN and NR wave amplitudes is roughly constant over
the course of the evolution, and therefore an equally good matching
between PN and NR waves should be possible with far less numerical
cycles. In particular, we performed a simulation that completes only 4.5
orbits before merger, and expect that this could be matched to PN 
waveforms by a procedure like that discussed in \cite{Ajith:2007qp,Ajith:2007kx} or 
\cite{Pan:2007nw} just as well as a simulation that models many more
cycles. We therefore conclude that, with the level of numerical accuracy 
that we can achieve, only about 4.5 orbits need be simulated for a 
PN/NR matching of the same accuracy. Whether these relatively modest
requirements for numerical waveforms carry over to the cases of unequal-mass
and nonspinning binaries will be the subject of future work.

For gravitational-wave detection we expect that such hybrid waveforms will
be acceptable. However, for parameter estimation the issue of the discrepancy 
between the amplitude of PN and NR waveforms may have to be addressed. 
Modeling the amplitude at 2.5PN order gives agreement within numerical error
between PN and NR waves up to about 11 cycles before merger; at present we
suggest that the best matching can be done with $>11$ cycles (5.5 orbits) of
numerical simulation. The cases where the current level of phase and amplitude
accuracy are expected to be adequate for various data-analysis applications
will also be explored in future work.

Comparing with evolutions of the PN equations of motion in the ADMTT gauge, we
find that the orbital motion seen in the numerical evolutions agrees extremely 
well up to a coordinate
separation of about $D = 8M$. This surprising agreement not
only suggests that the PN dynamics accurately models the full physical
dynamics up to about three orbits before merger, but that the numerical gauge
remains close to the ADMTT gauge up to that time. In addition, the gauge
dynamics and emission of junk radiation at the beginning of the simulation
do not noticeably change either the dynamics or the gauge; after about one
orbit the NR dynamics matches up again with the ADMTT PN dynamics.

{\it Note:} While this article was undergoing peer review, the
Caltech/Cornell group completed a detailed PN/NR comparison that covers 30
gravitational-wave cycles (15 orbits) before merger with high numerical
accuracy in their numerical waveforms \cite{Boyle:2007ft}. Where comparable,
their results confirm those in this paper; a comparison between our results
and theirs' is provided in their paper.

\acknowledgments
We thank Parameswaran Ajith, Duncan Brown and Mark Scheel for 
alerting us to errors in old versions of the TaylorT1 routines in LAL, which
were used to produce the results in an earlier draft of this paper. Since that
draft was written, Duncan Brown has committed to the LAL cvs repository bug fixes
to the relevant LAL routines. The subsequent changes in the TaylorT1 3.5PN
waveform phase changed the accumulated phase disagreement in
Figure~\ref{fig:PhaseDisagreement} from 0.15 radians to the current value of
0.8 radians. Prompted by these changes and by suggestions from the anonymous 
journal referee, 
we extended our analysis. However, the main conclusions from the first draft 
are unchanged.

We also thank Parameswaran Ajith and Alicia Sintes
for helpful discussions regarding the use of the LAL library,
Alessandra Buonanno and Achamveedu Gopakumar for clarifications 
regarding the TaylorT3 approximant, and Mark Scheel and Harald Pfeiffer for
feedback regarding our matching techniques. 
This work was supported in part by
DFG grant SFB/Transregio~7 ``Gravitational Wave Astronomy''.
We thank the DEISA Consortium (co-funded by the EU, FP6 project
508830), for support within the DEISA Extreme Computing Initiative
(www.deisa.org).
Computations were performed at LRZ Munich and the Doppler and Kepler
clusters at the Institute of Theoretical Physics of the University of Jena.
M.H.~and S.H.~
gratefully acknowledge hospitality of the University of the Balearic Islands.

\bibliography{refs}

\begin{thebibliography}{53}
\expandafter\ifx\csname natexlab\endcsname\relax\def\natexlab#1{#1}\fi
\expandafter\ifx\csname bibnamefont\endcsname\relax
  \def\bibnamefont#1{#1}\fi
\expandafter\ifx\csname bibfnamefont\endcsname\relax
  \def\bibfnamefont#1{#1}\fi
\expandafter\ifx\csname citenamefont\endcsname\relax
  \def\citenamefont#1{#1}\fi
\expandafter\ifx\csname url\endcsname\relax
  \def\url#1{\texttt{#1}}\fi
\expandafter\ifx\csname urlprefix\endcsname\relax\def\urlprefix{URL }\fi
\providecommand{\bibinfo}[2]{#2}
\providecommand{\eprint}[2][]{\url{#2}}

\bibitem[{\citenamefont{{Waldman (for the LIGO Scientific
  Collaboration)}}(2006)}]{Waldman06}
\bibinfo{author}{\bibfnamefont{S.}~\bibnamefont{{Waldman (for the LIGO
  Scientific Collaboration)}}}, \bibinfo{journal}{{Class. Quantum Grav. 23
  (2006) S653 -- S660}}  (\bibinfo{year}{2006}).

\bibitem[{\citenamefont{{Hild (for the LIGO Scientific
  Collaboration)}}(2006)}]{GEOStatus:2006}
\bibinfo{author}{\bibfnamefont{S.}~\bibnamefont{{Hild (for the LIGO Scientific
  Collaboration)}}}, \bibinfo{journal}{Class. Quantum Grav.}
  \textbf{\bibinfo{volume}{23}}, \bibinfo{pages}{S643} (\bibinfo{year}{2006}).

\bibitem[{\citenamefont{Acernese et~al.}(2006)}]{Acernese2006}
\bibinfo{author}{\bibfnamefont{F.}~\bibnamefont{Acernese}}
  \bibnamefont{et~al.}, \bibinfo{journal}{Class. Quantum Grav.}
  \textbf{\bibinfo{volume}{23}}, \bibinfo{pages}{S635} (\bibinfo{year}{2006}).

\bibitem[{\citenamefont{Buonanno et~al.}(2003)\citenamefont{Buonanno, Chen, and
  Vallisneri}}]{Buonanno03a}
\bibinfo{author}{\bibfnamefont{A.}~\bibnamefont{Buonanno}},
  \bibinfo{author}{\bibfnamefont{Y.}~\bibnamefont{Chen}}, \bibnamefont{and}
  \bibinfo{author}{\bibfnamefont{M.}~\bibnamefont{Vallisneri}},
  \bibinfo{journal}{Phys. Rev. D} \textbf{\bibinfo{volume}{67}},
  \bibinfo{pages}{024016} (\bibinfo{year}{2003}), \bibinfo{note}{erratum-ibid.
  D 74:029903 (2006)}, \eprint{gr-qc/0205122}.

\bibitem[{LAL()}]{LAL}
\emph{\bibinfo{title}{Lsc algorithm library {(LAL)}}},
  \urlprefix\url{http://www.lsc-group.phys.uwm.edu/lal}.

\bibitem[{\citenamefont{Pretorius}(2005)}]{Pretorius:2005gq}
\bibinfo{author}{\bibfnamefont{F.}~\bibnamefont{Pretorius}},
  \bibinfo{journal}{Phys. Rev. Lett.} \textbf{\bibinfo{volume}{95}},
  \bibinfo{pages}{121101} (\bibinfo{year}{2005}), \eprint{gr-qc/0507014}.

\bibitem[{\citenamefont{Campanelli
  et~al.}(2006{\natexlab{a}})\citenamefont{Campanelli, Lousto, Marronetti, and
  Zlochower}}]{Campanelli:2005dd}
\bibinfo{author}{\bibfnamefont{M.}~\bibnamefont{Campanelli}},
  \bibinfo{author}{\bibfnamefont{C.~O.} \bibnamefont{Lousto}},
  \bibinfo{author}{\bibfnamefont{P.}~\bibnamefont{Marronetti}},
  \bibnamefont{and}
  \bibinfo{author}{\bibfnamefont{Y.}~\bibnamefont{Zlochower}},
  \bibinfo{journal}{Phys. Rev. Lett.} \textbf{\bibinfo{volume}{96}},
  \bibinfo{pages}{111101} (\bibinfo{year}{2006}{\natexlab{a}}),
  \eprint{gr-qc/0511048}.

\bibitem[{\citenamefont{Baker et~al.}(2006{\natexlab{a}})\citenamefont{Baker,
  Centrella, Choi, Koppitz, and van Meter}}]{Baker05a}
\bibinfo{author}{\bibfnamefont{J.~G.} \bibnamefont{Baker}},
  \bibinfo{author}{\bibfnamefont{J.}~\bibnamefont{Centrella}},
  \bibinfo{author}{\bibfnamefont{D.-I.} \bibnamefont{Choi}},
  \bibinfo{author}{\bibfnamefont{M.}~\bibnamefont{Koppitz}}, \bibnamefont{and}
  \bibinfo{author}{\bibfnamefont{J.}~\bibnamefont{van Meter}},
  \bibinfo{journal}{Phys. Rev. Lett.} \textbf{\bibinfo{volume}{96}},
  \bibinfo{pages}{111102} (\bibinfo{year}{2006}{\natexlab{a}}),
  \eprint{gr-qc/0511103}.

\bibitem[{\citenamefont{Pretorius}(2006)}]{Pretorius:2006tp}
\bibinfo{author}{\bibfnamefont{F.}~\bibnamefont{Pretorius}},
  \bibinfo{journal}{Class. Quantum Grav.} \textbf{\bibinfo{volume}{23}},
  \bibinfo{pages}{S529} (\bibinfo{year}{2006}), \eprint{gr-qc/0602115}.

\bibitem[{\citenamefont{Campanelli
  et~al.}(2006{\natexlab{b}})\citenamefont{Campanelli, Lousto, and
  Zlochower}}]{Campanelli:2006gf}
\bibinfo{author}{\bibfnamefont{M.}~\bibnamefont{Campanelli}},
  \bibinfo{author}{\bibfnamefont{C.~O.} \bibnamefont{Lousto}},
  \bibnamefont{and}
  \bibinfo{author}{\bibfnamefont{Y.}~\bibnamefont{Zlochower}},
  \bibinfo{journal}{Phys. Rev. D} \textbf{\bibinfo{volume}{73}},
  \bibinfo{pages}{061501(R)} (\bibinfo{year}{2006}{\natexlab{b}}),
  \eprint{gr-qc/0601091}.

\bibitem[{\citenamefont{Baker et~al.}(2006{\natexlab{b}})\citenamefont{Baker,
  Centrella, Choi, Koppitz, and van Meter}}]{Baker:2006yw}
\bibinfo{author}{\bibfnamefont{J.~G.} \bibnamefont{Baker}},
  \bibinfo{author}{\bibfnamefont{J.}~\bibnamefont{Centrella}},
  \bibinfo{author}{\bibfnamefont{D.-I.} \bibnamefont{Choi}},
  \bibinfo{author}{\bibfnamefont{M.}~\bibnamefont{Koppitz}}, \bibnamefont{and}
  \bibinfo{author}{\bibfnamefont{J.}~\bibnamefont{van Meter}},
  \bibinfo{journal}{Phys. Rev. D} \textbf{\bibinfo{volume}{73}},
  \bibinfo{pages}{104002} (\bibinfo{year}{2006}{\natexlab{b}}),
  \eprint{gr-qc/0602026}.

\bibitem[{\citenamefont{Sperhake}(2006)}]{Sperhake2006}
\bibinfo{author}{\bibfnamefont{U.}~\bibnamefont{Sperhake}}
  (\bibinfo{year}{2006}), \bibinfo{note}{gr-qc/0606079}.

\bibitem[{\citenamefont{Br{\"u}gmann et~al.}(2006)\citenamefont{Br{\"u}gmann,
  Gonz{\'a}lez, Hannam, Husa, Sperhake, and Tichy}}]{Bruegmann:2006at}
\bibinfo{author}{\bibfnamefont{B.}~\bibnamefont{Br{\"u}gmann}},
  \bibinfo{author}{\bibfnamefont{J.~A.} \bibnamefont{Gonz{\'a}lez}},
  \bibinfo{author}{\bibfnamefont{M.}~\bibnamefont{Hannam}},
  \bibinfo{author}{\bibfnamefont{S.}~\bibnamefont{Husa}},
  \bibinfo{author}{\bibfnamefont{U.}~\bibnamefont{Sperhake}}, \bibnamefont{and}
  \bibinfo{author}{\bibfnamefont{W.}~\bibnamefont{Tichy}}
  (\bibinfo{year}{2006}), \bibinfo{note}{gr-qc/0610128}.

\bibitem[{\citenamefont{Baker et~al.}(2006{\natexlab{c}})\citenamefont{Baker,
  McWilliams, van Meter, Centrella, Choi, Kelly, and Koppitz}}]{Baker:2006ls}
\bibinfo{author}{\bibfnamefont{J.~G.} \bibnamefont{Baker}},
  \bibinfo{author}{\bibfnamefont{S.~T.} \bibnamefont{McWilliams}},
  \bibinfo{author}{\bibfnamefont{J.~R.} \bibnamefont{van Meter}},
  \bibinfo{author}{\bibfnamefont{J.}~\bibnamefont{Centrella}},
  \bibinfo{author}{\bibfnamefont{D.-I.} \bibnamefont{Choi}},
  \bibinfo{author}{\bibfnamefont{B.~J.} \bibnamefont{Kelly}}, \bibnamefont{and}
  \bibinfo{author}{\bibfnamefont{M.}~\bibnamefont{Koppitz}}
  (\bibinfo{year}{2006}{\natexlab{c}}), \bibinfo{note}{gr-qc/0612117},
  \eprint{gr-qc/0612117}.

\bibitem[{\citenamefont{Baker et~al.}(2006{\natexlab{d}})\citenamefont{Baker,
  van Meter, McWilliams, Centrella, and Kelly}}]{Baker:2006ha}
\bibinfo{author}{\bibfnamefont{J.~G.} \bibnamefont{Baker}},
  \bibinfo{author}{\bibfnamefont{J.~R.} \bibnamefont{van Meter}},
  \bibinfo{author}{\bibfnamefont{S.~T.} \bibnamefont{McWilliams}},
  \bibinfo{author}{\bibfnamefont{J.}~\bibnamefont{Centrella}},
  \bibnamefont{and} \bibinfo{author}{\bibfnamefont{B.~J.} \bibnamefont{Kelly}}
  (\bibinfo{year}{2006}{\natexlab{d}}), \eprint{gr-qc/0612024}.

\bibitem[{\citenamefont{Pfeiffer et~al.}(2007)\citenamefont{Pfeiffer, Brown,
  Kidder, Lindblom, Lovelance, and Scheel}}]{Pfeiffer:2007}
\bibinfo{author}{\bibfnamefont{H.~P.} \bibnamefont{Pfeiffer}},
  \bibinfo{author}{\bibfnamefont{D.}~\bibnamefont{Brown}},
  \bibinfo{author}{\bibfnamefont{L.~E.} \bibnamefont{Kidder}},
  \bibinfo{author}{\bibfnamefont{L.}~\bibnamefont{Lindblom}},
  \bibinfo{author}{\bibfnamefont{G.}~\bibnamefont{Lovelance}},
  \bibnamefont{and} \bibinfo{author}{\bibfnamefont{M.~A.} \bibnamefont{Scheel}}
  (\bibinfo{year}{2007}), \eprint{gr-qc/0702106}.

\bibitem[{\citenamefont{Scheel et~al.}(2006)\citenamefont{Scheel, Pfeiffer,
  Lindblom, Kidder, Rinne, and Teukolsky}}]{Scheel-etal-2006:dual-frame}
\bibinfo{author}{\bibfnamefont{M.~A.} \bibnamefont{Scheel}},
  \bibinfo{author}{\bibfnamefont{H.~P.} \bibnamefont{Pfeiffer}},
  \bibinfo{author}{\bibfnamefont{L.}~\bibnamefont{Lindblom}},
  \bibinfo{author}{\bibfnamefont{L.~E.} \bibnamefont{Kidder}},
  \bibinfo{author}{\bibfnamefont{O.}~\bibnamefont{Rinne}}, \bibnamefont{and}
  \bibinfo{author}{\bibfnamefont{S.~A.} \bibnamefont{Teukolsky}},
  \bibinfo{journal}{Phys. Rev. D} \textbf{\bibinfo{volume}{74}},
  \bibinfo{pages}{104006} (\bibinfo{year}{2006}), \eprint{gr-qc/0607056}.

\bibitem[{\citenamefont{Husa et~al.}(2007{\natexlab{a}})\citenamefont{Husa,
  Gonz{\'a}lez, Hannam, Br{\"u}gmann, and Sperhake}}]{Husa2007a}
\bibinfo{author}{\bibfnamefont{S.}~\bibnamefont{Husa}},
  \bibinfo{author}{\bibfnamefont{J.~A.} \bibnamefont{Gonz{\'a}lez}},
  \bibinfo{author}{\bibfnamefont{M.}~\bibnamefont{Hannam}},
  \bibinfo{author}{\bibfnamefont{B.}~\bibnamefont{Br{\"u}gmann}},
  \bibnamefont{and} \bibinfo{author}{\bibfnamefont{U.}~\bibnamefont{Sperhake}}
  (\bibinfo{year}{2007}{\natexlab{a}}), \eprint{arXiv:0706.0740 [gr-qc]}.

\bibitem[{\citenamefont{Buonanno
  et~al.}(2006{\natexlab{a}})\citenamefont{Buonanno, Cook, and
  Pretorius}}]{Buonanno06imr}
\bibinfo{author}{\bibfnamefont{A.}~\bibnamefont{Buonanno}},
  \bibinfo{author}{\bibfnamefont{G.~B.} \bibnamefont{Cook}}, \bibnamefont{and}
  \bibinfo{author}{\bibfnamefont{F.}~\bibnamefont{Pretorius}}
  (\bibinfo{year}{2006}{\natexlab{a}}), \bibinfo{note}{gr-qc/0610122},
  \eprint{gr-qc/0610122}.

\bibitem[{\citenamefont{Berti et~al.}(2007)}]{Berti:2007fi}
\bibinfo{author}{\bibfnamefont{E.}~\bibnamefont{Berti}} \bibnamefont{et~al.}
  (\bibinfo{year}{2007}), \bibinfo{note}{gr-qc/0703053},
  \eprint{gr-qc/0703053}.

\bibitem[{\citenamefont{Pan et~al.}(2007)}]{Pan:2007nw}
\bibinfo{author}{\bibfnamefont{Y.}~\bibnamefont{Pan}} \bibnamefont{et~al.}
  (\bibinfo{year}{2007}), \bibinfo{note}{arXiv:0704.1964 [gr-qc]},
  \eprint{arXiv:0704.1964 [gr-qc]}.

\bibitem[{\citenamefont{Buonanno et~al.}(2007)\citenamefont{Buonanno, Pan,
  Baker, Centrella, Kelly, McWilliams, and {van Meter}}}]{Buonanno:2007pf}
\bibinfo{author}{\bibfnamefont{A.}~\bibnamefont{Buonanno}},
  \bibinfo{author}{\bibfnamefont{Y.}~\bibnamefont{Pan}},
  \bibinfo{author}{\bibfnamefont{J.~G.} \bibnamefont{Baker}},
  \bibinfo{author}{\bibfnamefont{J.}~\bibnamefont{Centrella}},
  \bibinfo{author}{\bibfnamefont{B.~J.} \bibnamefont{Kelly}},
  \bibinfo{author}{\bibfnamefont{S.~T.} \bibnamefont{McWilliams}},
  \bibnamefont{and} \bibinfo{author}{\bibfnamefont{J.~R.} \bibnamefont{{van
  Meter}}} (\bibinfo{year}{2007}), \eprint{arXiv:0706.3732 [gr-qc]}.

\bibitem[{\citenamefont{Vaishnav et~al.}(2007)\citenamefont{Vaishnav, Hinder,
  Herrmann, and Shoemaker}}]{Vaishnav:2007nm}
\bibinfo{author}{\bibfnamefont{B.}~\bibnamefont{Vaishnav}},
  \bibinfo{author}{\bibfnamefont{I.}~\bibnamefont{Hinder}},
  \bibinfo{author}{\bibfnamefont{F.}~\bibnamefont{Herrmann}}, \bibnamefont{and}
  \bibinfo{author}{\bibfnamefont{D.}~\bibnamefont{Shoemaker}}
  (\bibinfo{year}{2007}), \eprint{arXiv:0705.3829 [gr-qc]}.

\bibitem[{\citenamefont{{Ajith} et~al.}(2007)\citenamefont{{Ajith}, {Babak},
  {Chen}, {Hewitson}, {Krishnan}, {Whelan}, {Br\"ugmann}, {Diener},
  {Gonz\'alez}, {Hannam} et~al.}}]{Ajith:2007qp}
\bibinfo{author}{\bibfnamefont{P.}~\bibnamefont{{Ajith}}},
  \bibinfo{author}{\bibfnamefont{S.}~\bibnamefont{{Babak}}},
  \bibinfo{author}{\bibfnamefont{Y.}~\bibnamefont{{Chen}}},
  \bibinfo{author}{\bibfnamefont{M.}~\bibnamefont{{Hewitson}}},
  \bibinfo{author}{\bibfnamefont{B.}~\bibnamefont{{Krishnan}}},
  \bibinfo{author}{\bibfnamefont{J.~T.} \bibnamefont{{Whelan}}},
  \bibinfo{author}{\bibfnamefont{B.}~\bibnamefont{{Br\"ugmann}}},
  \bibinfo{author}{\bibfnamefont{P.}~\bibnamefont{{Diener}}},
  \bibinfo{author}{\bibfnamefont{J.}~\bibnamefont{{Gonz\'alez}}},
  \bibinfo{author}{\bibfnamefont{M.}~\bibnamefont{{Hannam}}},
  \bibnamefont{et~al.}, \bibinfo{journal}{Class. Quantum Grav.}
  \textbf{\bibinfo{volume}{24}}, \bibinfo{pages}{S689} (\bibinfo{year}{2007}),
  \eprint{arXiv:0704.3764 [gr-qc]}.

\bibitem[{\citenamefont{Ajith et~al.}(2007)}]{Ajith:2007kx}
\bibinfo{author}{\bibfnamefont{P.}~\bibnamefont{Ajith}} \bibnamefont{et~al.}
  (\bibinfo{year}{2007}), \eprint{arXiv:0710.2335 [gr-qc]}.

\bibitem[{\citenamefont{Buonanno
  et~al.}(2006{\natexlab{b}})\citenamefont{Buonanno, Chen, and
  Damour}}]{Buonanno:2005xu}
\bibinfo{author}{\bibfnamefont{A.}~\bibnamefont{Buonanno}},
  \bibinfo{author}{\bibfnamefont{Y.}~\bibnamefont{Chen}}, \bibnamefont{and}
  \bibinfo{author}{\bibfnamefont{T.}~\bibnamefont{Damour}},
  \bibinfo{journal}{Phys. Rev.} \textbf{\bibinfo{volume}{D74}},
  \bibinfo{pages}{104005} (\bibinfo{year}{2006}{\natexlab{b}}),
  \eprint{gr-qc/0508067}.

\bibitem[{\citenamefont{Husa et~al.}(2007{\natexlab{b}})\citenamefont{Husa,
  Hannam, Gonz{\'a}lez, Sperhake, and Br{\"u}gmann}}]{Husa:2007ec}
\bibinfo{author}{\bibfnamefont{S.}~\bibnamefont{Husa}},
  \bibinfo{author}{\bibfnamefont{M.}~\bibnamefont{Hannam}},
  \bibinfo{author}{\bibfnamefont{J.~A.} \bibnamefont{Gonz{\'a}lez}},
  \bibinfo{author}{\bibfnamefont{U.}~\bibnamefont{Sperhake}}, \bibnamefont{and}
  \bibinfo{author}{\bibfnamefont{B.}~\bibnamefont{Br{\"u}gmann}}
  (\bibinfo{year}{2007}{\natexlab{b}}), \eprint{arXiv:0706.0904 [gr-qc]}.

\bibitem[{\citenamefont{Br{\"u}gmann et~al.}(2004)\citenamefont{Br{\"u}gmann,
  Tichy, and Jansen}}]{Bruegmann2004}
\bibinfo{author}{\bibfnamefont{B.}~\bibnamefont{Br{\"u}gmann}},
  \bibinfo{author}{\bibfnamefont{W.}~\bibnamefont{Tichy}}, \bibnamefont{and}
  \bibinfo{author}{\bibfnamefont{N.}~\bibnamefont{Jansen}},
  \bibinfo{journal}{Phys. Rev. Lett.} \textbf{\bibinfo{volume}{92}},
  \bibinfo{pages}{211101} (\bibinfo{year}{2004}),
  \bibinfo{note}{gr-qc/0312112}.

\bibitem[{\citenamefont{Brandt and Br{\"u}gmann}(1997)}]{Brandt97b}
\bibinfo{author}{\bibfnamefont{S.}~\bibnamefont{Brandt}} \bibnamefont{and}
  \bibinfo{author}{\bibfnamefont{B.}~\bibnamefont{Br{\"u}gmann}},
  \bibinfo{journal}{Phys. Rev. Lett.} \textbf{\bibinfo{volume}{78}},
  \bibinfo{pages}{3606} (\bibinfo{year}{1997}), \eprint{gr-qc/9703066}.

\bibitem[{\citenamefont{Bowen and York}(1980)}]{Bowen80}
\bibinfo{author}{\bibfnamefont{J.~M.} \bibnamefont{Bowen}} \bibnamefont{and}
  \bibinfo{author}{\bibfnamefont{J.~W.} \bibnamefont{York}},
  \bibinfo{journal}{Phys. Rev. D} \textbf{\bibinfo{volume}{21}},
  \bibinfo{pages}{2047} (\bibinfo{year}{1980}).

\bibitem[{\citenamefont{Ansorg et~al.}(2004)\citenamefont{Ansorg, Br{\"u}gmann,
  and Tichy}}]{Ansorg:2004ds}
\bibinfo{author}{\bibfnamefont{M.}~\bibnamefont{Ansorg}},
  \bibinfo{author}{\bibfnamefont{B.}~\bibnamefont{Br{\"u}gmann}},
  \bibnamefont{and} \bibinfo{author}{\bibfnamefont{W.}~\bibnamefont{Tichy}},
  \bibinfo{journal}{Phys. Rev. D} \textbf{\bibinfo{volume}{70}},
  \bibinfo{pages}{064011} (\bibinfo{year}{2004}), \eprint{gr-qc/0404056}.

\bibitem[{\citenamefont{Campanelli
  et~al.}(2006{\natexlab{c}})\citenamefont{Campanelli, Lousto, Marronetti, and
  Zlochower}}]{Campanelli2006}
\bibinfo{author}{\bibfnamefont{M.}~\bibnamefont{Campanelli}},
  \bibinfo{author}{\bibfnamefont{C.~O.} \bibnamefont{Lousto}},
  \bibinfo{author}{\bibfnamefont{P.}~\bibnamefont{Marronetti}},
  \bibnamefont{and}
  \bibinfo{author}{\bibfnamefont{Y.}~\bibnamefont{Zlochower}},
  \bibinfo{journal}{Phys. Rev. Lett.} \textbf{\bibinfo{volume}{96}},
  \bibinfo{pages}{111101} (\bibinfo{year}{2006}{\natexlab{c}}),
  \bibinfo{note}{gr-qc/0511048}.

\bibitem[{\citenamefont{Baker et~al.}(2006{\natexlab{e}})\citenamefont{Baker,
  Centrella, Choi, Koppitz, and van Meter}}]{Baker2006}
\bibinfo{author}{\bibfnamefont{J.~G.} \bibnamefont{Baker}},
  \bibinfo{author}{\bibfnamefont{J.}~\bibnamefont{Centrella}},
  \bibinfo{author}{\bibfnamefont{D.-I.} \bibnamefont{Choi}},
  \bibinfo{author}{\bibfnamefont{M.}~\bibnamefont{Koppitz}}, \bibnamefont{and}
  \bibinfo{author}{\bibfnamefont{J.}~\bibnamefont{van Meter}},
  \bibinfo{journal}{Phys. Rev. Lett.} \textbf{\bibinfo{volume}{96}},
  \bibinfo{pages}{111102} (\bibinfo{year}{2006}{\natexlab{e}}),
  \bibinfo{note}{gr-qc/0511103}.

\bibitem[{\citenamefont{Shibata and Nakamura}(1995)}]{Shibata95}
\bibinfo{author}{\bibfnamefont{M.}~\bibnamefont{Shibata}} \bibnamefont{and}
  \bibinfo{author}{\bibfnamefont{T.}~\bibnamefont{Nakamura}},
  \bibinfo{journal}{Phys. Rev. D} \textbf{\bibinfo{volume}{52}},
  \bibinfo{pages}{5428} (\bibinfo{year}{1995}).

\bibitem[{\citenamefont{Baumgarte and Shapiro}(1999)}]{Baumgarte99}
\bibinfo{author}{\bibfnamefont{T.~W.} \bibnamefont{Baumgarte}}
  \bibnamefont{and} \bibinfo{author}{\bibfnamefont{S.~L.}
  \bibnamefont{Shapiro}}, \bibinfo{journal}{Phys. Rev. D}
  \textbf{\bibinfo{volume}{59}}, \bibinfo{pages}{024007}
  (\bibinfo{year}{1999}), \eprint{gr-qc/9810065}.

\bibitem[{\citenamefont{York}(1979)}]{York79}
\bibinfo{author}{\bibfnamefont{J.~W.} \bibnamefont{York}}, in
  \emph{\bibinfo{booktitle}{Sources of gravitational radiation}}, edited by
  \bibinfo{editor}{\bibfnamefont{L.~L.} \bibnamefont{Smarr}}
  (\bibinfo{publisher}{Cambridge University Press},
  \bibinfo{address}{Cambridge, UK}, \bibinfo{year}{1979}), pp.
  \bibinfo{pages}{83--126}, ISBN \bibinfo{isbn}{0-521-22778-X}.

\bibitem[{\citenamefont{Tichy and Br{\"u}gmann}(2004)}]{Tichy:2003qi}
\bibinfo{author}{\bibfnamefont{W.}~\bibnamefont{Tichy}} \bibnamefont{and}
  \bibinfo{author}{\bibfnamefont{B.}~\bibnamefont{Br{\"u}gmann}},
  \bibinfo{journal}{Phys. Rev. D} \textbf{\bibinfo{volume}{69}},
  \bibinfo{pages}{024006} (\bibinfo{year}{2004}), \eprint{gr-qc/0307027}.

\bibitem[{\citenamefont{Cook and York}(1990)}]{Cook90a}
\bibinfo{author}{\bibfnamefont{G.~B.} \bibnamefont{Cook}} \bibnamefont{and}
  \bibinfo{author}{\bibfnamefont{J.~W.} \bibnamefont{York},
  \bibfnamefont{Jr.}}, \bibinfo{journal}{Phys. Rev. D}
  \textbf{\bibinfo{volume}{41}}, \bibinfo{pages}{1077} (\bibinfo{year}{1990}),
  \urlprefix\url{http://link.aps.org/abstract/PRD/v41/p1077}.

\bibitem[{\citenamefont{Cook}(1990)}]{Cook90}
\bibinfo{author}{\bibfnamefont{G.~B.} \bibnamefont{Cook}}, Ph.D. thesis,
  \bibinfo{school}{University of North Carolina at Chapel Hill},
  \bibinfo{address}{Chapel Hill, North Carolina} (\bibinfo{year}{1990}).

\bibitem[{\citenamefont{Campanelli
  et~al.}(2006{\natexlab{d}})\citenamefont{Campanelli, Lousto, and
  Zlochower}}]{Campanelli:2006vp}
\bibinfo{author}{\bibfnamefont{M.}~\bibnamefont{Campanelli}},
  \bibinfo{author}{\bibfnamefont{C.~O.} \bibnamefont{Lousto}},
  \bibnamefont{and}
  \bibinfo{author}{\bibfnamefont{Y.}~\bibnamefont{Zlochower}},
  \bibinfo{journal}{Phys. Rev. D} \textbf{\bibinfo{volume}{74}},
  \bibinfo{pages}{084023} (\bibinfo{year}{2006}{\natexlab{d}}),
  \eprint{gr-qc/0608275}.

\bibitem[{\citenamefont{Damour et~al.}(2001)\citenamefont{Damour, Iyer, and
  Sathyaprakash}}]{Damour00a}
\bibinfo{author}{\bibfnamefont{T.}~\bibnamefont{Damour}},
  \bibinfo{author}{\bibfnamefont{B.~R.} \bibnamefont{Iyer}}, \bibnamefont{and}
  \bibinfo{author}{\bibfnamefont{B.~S.} \bibnamefont{Sathyaprakash}},
  \bibinfo{journal}{Phys. Rev. D} \textbf{\bibinfo{volume}{63}},
  \bibinfo{pages}{044023} (\bibinfo{year}{2001}), \bibinfo{note}{erratum Phys.
  Rev. D{\bf 72}, 029902 (2005).}, \eprint{gr-qc/0010009}.

\bibitem[{\citenamefont{Damour et~al.}(2002)\citenamefont{Damour, Iyer, and
  Sathyaprakash}}]{Damour:2002kr}
\bibinfo{author}{\bibfnamefont{T.}~\bibnamefont{Damour}},
  \bibinfo{author}{\bibfnamefont{B.~R.} \bibnamefont{Iyer}}, \bibnamefont{and}
  \bibinfo{author}{\bibfnamefont{B.~S.} \bibnamefont{Sathyaprakash}},
  \bibinfo{journal}{Phys. Rev.} \textbf{\bibinfo{volume}{D66}},
  \bibinfo{pages}{027502} (\bibinfo{year}{2002}), \bibinfo{note}{erratum Phys.
  Rev. D{\bf 72}, 029901 (2005).}, \eprint{gr-qc/0207021}.

\bibitem[{\citenamefont{Arun et~al.}(2004)\citenamefont{Arun, Blanchet, Iyer,
  and Qusailah}}]{Arun04}
\bibinfo{author}{\bibfnamefont{K.}~\bibnamefont{Arun}},
  \bibinfo{author}{\bibfnamefont{L.}~\bibnamefont{Blanchet}},
  \bibinfo{author}{\bibfnamefont{B.~R.} \bibnamefont{Iyer}}, \bibnamefont{and}
  \bibinfo{author}{\bibfnamefont{M.~S.~S.} \bibnamefont{Qusailah}},
  \bibinfo{journal}{Class. Quantum Grav.} \textbf{\bibinfo{volume}{21}},
  \bibinfo{pages}{3771} (\bibinfo{year}{2004}), \bibinfo{note}{erratum-ibid.
  22, 3115, (2005)}.

\bibitem[{\citenamefont{Blanchet et~al.}(2004)\citenamefont{Blanchet, Damour,
  Esposito-Farese, and Iyer}}]{Blanchet:2004ek}
\bibinfo{author}{\bibfnamefont{L.}~\bibnamefont{Blanchet}},
  \bibinfo{author}{\bibfnamefont{T.}~\bibnamefont{Damour}},
  \bibinfo{author}{\bibfnamefont{G.}~\bibnamefont{Esposito-Farese}},
  \bibnamefont{and} \bibinfo{author}{\bibfnamefont{B.~R.} \bibnamefont{Iyer}},
  \bibinfo{journal}{Phys. Rev. Lett.} \textbf{\bibinfo{volume}{93}},
  \bibinfo{pages}{091101} (\bibinfo{year}{2004}), \eprint{gr-qc/0406012}.

\bibitem[{\citenamefont{Blanchet et~al.}(2002)\citenamefont{Blanchet, Faye,
  Iyer, and Joguet}}]{Blanchet:2001ax}
\bibinfo{author}{\bibfnamefont{L.}~\bibnamefont{Blanchet}},
  \bibinfo{author}{\bibfnamefont{G.}~\bibnamefont{Faye}},
  \bibinfo{author}{\bibfnamefont{B.~R.} \bibnamefont{Iyer}}, \bibnamefont{and}
  \bibinfo{author}{\bibfnamefont{B.}~\bibnamefont{Joguet}},
  \bibinfo{journal}{Phys. Rev. D} \textbf{\bibinfo{volume}{65}},
  \bibinfo{pages}{061501} (\bibinfo{year}{2002}), \bibinfo{note}{erratum Phys.
  Rev. D{\bf 71}, 129902 (2005).}, \eprint{gr-qc/0105099}.

\bibitem[{\citenamefont{Sundararajan et~al.}(2007)\citenamefont{Sundararajan,
  Khanna, and Hughes}}]{Sundararajan:2007jg}
\bibinfo{author}{\bibfnamefont{P.~A.} \bibnamefont{Sundararajan}},
  \bibinfo{author}{\bibfnamefont{G.}~\bibnamefont{Khanna}}, \bibnamefont{and}
  \bibinfo{author}{\bibfnamefont{S.~A.} \bibnamefont{Hughes}}
  (\bibinfo{year}{2007}), \eprint{gr-qc/0703028}.

\bibitem[{\citenamefont{Krishnan et~al.}(2007)\citenamefont{Krishnan, Lousto,
  and Zlochower}}]{Krishnan:2007pu}
\bibinfo{author}{\bibfnamefont{B.}~\bibnamefont{Krishnan}},
  \bibinfo{author}{\bibfnamefont{C.~O.} \bibnamefont{Lousto}},
  \bibnamefont{and} \bibinfo{author}{\bibfnamefont{Y.}~\bibnamefont{Zlochower}}
  (\bibinfo{year}{2007}), \eprint{arXiv:0707.0876 [gr-qc]}.

\bibitem[{\citenamefont{Lousto and Zlochower}(2007)}]{Lousto:2007db}
\bibinfo{author}{\bibfnamefont{C.~O.} \bibnamefont{Lousto}} \bibnamefont{and}
  \bibinfo{author}{\bibfnamefont{Y.}~\bibnamefont{Zlochower}}
  (\bibinfo{year}{2007}), \eprint{arXiv:0708.4048 [gr-qc]}.

\bibitem[{\citenamefont{Fiske et~al.}(2005)\citenamefont{Fiske, Baker, van
  Meter, Choi, and Centrella}}]{Fiske05}
\bibinfo{author}{\bibfnamefont{D.~R.} \bibnamefont{Fiske}},
  \bibinfo{author}{\bibfnamefont{J.~G.} \bibnamefont{Baker}},
  \bibinfo{author}{\bibfnamefont{J.~R.} \bibnamefont{van Meter}},
  \bibinfo{author}{\bibfnamefont{D.}~\bibnamefont{Choi}}, \bibnamefont{and}
  \bibinfo{author}{\bibfnamefont{J.~M.} \bibnamefont{Centrella}},
  \bibinfo{journal}{Phys. Rev. D} \textbf{\bibinfo{volume}{71}},
  \bibinfo{pages}{104036} (\bibinfo{year}{2005}),
  \bibinfo{note}{gr-qc/0503100}.

\bibitem[{\citenamefont{Br{\"u}gmann et~al.}(2007)\citenamefont{Br{\"u}gmann,
  Gonz{\'a}lez, Hannam, Husa, and Sperhake}}]{Bruegmann:2007a}
\bibinfo{author}{\bibfnamefont{B.}~\bibnamefont{Br{\"u}gmann}},
  \bibinfo{author}{\bibfnamefont{J.~A.} \bibnamefont{Gonz{\'a}lez}},
  \bibinfo{author}{\bibfnamefont{M.}~\bibnamefont{Hannam}},
  \bibinfo{author}{\bibfnamefont{S.}~\bibnamefont{Husa}}, \bibnamefont{and}
  \bibinfo{author}{\bibfnamefont{U.}~\bibnamefont{Sperhake}}
  (\bibinfo{year}{2007}), \bibinfo{note}{in preparation}.

\bibitem[{\citenamefont{Kidder}(1995)}]{Kidder1995}
\bibinfo{author}{\bibfnamefont{L.~E.} \bibnamefont{Kidder}},
  \bibinfo{journal}{Phys. Rev. D} \textbf{\bibinfo{volume}{52}},
  \bibinfo{pages}{821} (\bibinfo{year}{1995}).

\bibitem[{\citenamefont{Jaranowski and Sch{\"a}fer}(1998)}]{Jaranowski98a}
\bibinfo{author}{\bibfnamefont{P.}~\bibnamefont{Jaranowski}} \bibnamefont{and}
  \bibinfo{author}{\bibfnamefont{G.}~\bibnamefont{Sch{\"a}fer}},
  \bibinfo{journal}{Phys. Rev. D} \textbf{\bibinfo{volume}{57}},
  \bibinfo{pages}{7274} (\bibinfo{year}{1998}).

\bibitem[{\citenamefont{Boyle et~al.}(2007)}]{Boyle:2007ft}
\bibinfo{author}{\bibfnamefont{M.}~\bibnamefont{Boyle}} \bibnamefont{et~al.}
  (\bibinfo{year}{2007}), \eprint{arXiv:0710.0158 [gr-qc]}.

\end{thebibliography}

\end{document}